\documentclass{rsauthor}
\usepackage{amsmath}
\usepackage{amssymb}
\usepackage{graphicx}
\usepackage{subfigure}
\usepackage{bm}
\usepackage{nomencl}

\usepackage{comment} 
\usepackage{natbib}
\usepackage{pdfpages} 

\newcommand{\mbf}{\bm}
\newcommand{\gras}{\bm}
\newcommand{\gradient}{\gras{\nabla}}  

\newcommand{\del}{\partial}                             
\newcommand{\Deli}[2]{\frac{D #1}{D #2}}          

\newcommand{\intx}[4]{\int_{#1}^{#2} #3\, d#4}

\newcommand{\intxInf}[2]{\intx{-\infty}{+\infty}{#1}{#2}}

\newcommand{\scalar}[2]{{\gras #1} \cdot \gras #2}

\newcommand{\erf}[1]{\text{erf}{\left(#1\right)}}

\newcommand{\column}[2]{ \begin{pmatrix} #1 \\ #2 \end{pmatrix} }
\newcommand{\mymatrix}[4]{ \begin{pmatrix} #1 & #2 \\ #3 & #4 \end{pmatrix} }

\jname{Proc. Roy. Soc. A.}
\makenomenclature

\title{Trailing edge noise theory for rotating blades in uniform flow}

\author{%
    S. Sinayoko$^{1, *}$
    \thanks{$^*$ Author for correspondance (s.sinayoko@eng.cam.ac.uk)},
  M. Kingan$^{2}$, %
  \quad A. Agarwal$^{3}$
}

\address{$^{1,3}$ Department of Engineering, University of Cambridge, \\Cambridge CB2 1PZ, United Kingdom \\%
$^{2}$ Institute of Sound and Vibration Research, University of Southampton, \\Southamption SO17 1BJ, United Kingdom}

\date{\today}

\begin{document}
\maketitle

\begin{abstract}%
TThis paper presents a new formulation for trailing edge noise radiation from rotating blades based on an analytical solution of the convective wave equation. 
It accounts for distributed loading and the effect of mean flow and spanwise wavenumber.
A commonly used theory due to Schlinker and Amiet~(\citeyear{Schlinker1981a}) predicts
trailing edge noise radiation from rotating blades.
However, different versions of the theory exist; it is not known which version is the correct one, and what the range
of validity of the theory is.
This paper addresses both questions by deriving Schlinker and Amiet's theory in a simple way and by comparing it to the new formulation, using model blade elements representative of
a wind turbine, a cooling fan and an aircraft propeller. 
The correct form of Schlinker and Amiet's theory~(\citeyear{Schlinker1981a}) is identified. 
It is valid at high enough frequency, i.e. for a Helmholtz number relative
to chord greater than one and a rotational frequency much smaller than the angular frequency of the noise sources. 
\end{abstract}

\section{Introduction}

Turbulent eddies convecting within the boundary layer of an aerofoil are scattered into sound at the trailing edge.
This, turbulent boundary layer trailing edge noise, is a major source of broadband noise for an aerofoil in a uniform flow.
It is the dominant noise source for large wind turbines~\citep{Oerlemans2007}.
For installed fans, trailing edge noise corresponds to the minimum achievable noise level~\citep{wright1976acoustic}. 

Although this paper focuses on trailing edge noise, rotors and propellers are subject to several other noise sources~\citep{Hubbard1991}. Some of these are broadband, for example leading edge noise \citep{Paterson1982, Homicz1974} due to upstream turbulence, tip vortex induced noise and stall noise~\citep{Moreau2009_AIAA}. Others are tonal, as in the case of: steady loading in the reference frame of the rotor (\citet{Gutin1948},~\citet{hanson_influence_1980}); periodic unsteady loading produced by stationary distortions in the flow. 

Trailing edge noise can be predicted in the time domain by solving the {Ffowcs-Williams} \& Hawkings (FWH) equation (\cite{FfowcsWilliams1969, MorfeySept.2007, Najafi-Yazdi2010}), as demonstrated by~\cite{Casper2004}. However, this type of prediction requires numerical differentiations and the calculation of retarded times that are computationally expensive. Most importantly, the roles of blade geometry and operating conditions become clearer with the frequency domain formulations that are the subject of this paper. Another key advantage of the frequency domain formulation is that it is more suited  for a statistical description of the noise sources.  

One of the most successful frequency domain formulations was developed by Amiet (\citeyear{Amiet1976, Amiet1978}). 
Assuming knowledge of the pressure fluctuations travelling towards the trailing edge, Amiet's formulation gives an analytical expression for the power spectral density in the far field. This analytical expression makes it efficient and attractive for fast turn-around applications. It has been applied to numerous applications including low speed fans~(\cite{Rozenberg2010}), helicopters~(\cite{Paterson1982,Amiet1986}) and wind turbines~(\cite{Glegg1987}). Although Amiet's approach was initially restricted to high frequencies relative to the chord ($k C > 1$), it has been extended to lower frequencies ($k C > 0. 1$) by~\cite{Roger2005}. 
\nomenclature{$k$}{Acoustic wavenumber $\omega/c_0$}%

The effect of rotation on trailing edge noise modelling was analysed in \cite{Amiet1977} and \cite{Schlinker1981a} for a rotor in a uniform flow.
The first step is to estimate the instantaneous power spectral density radiating from the blade, while it is located at a particular azimuthal angle around the hub.
Secondly, the power spectra are averaged in time.
The final expression involves a Doppler factor of the form $(\omega'/ \omega)^{a}$, where $\omega'$ is the source frequency and $\omega$ the frequency of the observed sound.
The exponent $a$ takes the value 1 in \cite{Amiet1976, Rozenberg2010}, 2 in \cite{Schlinker1981a} and -2 in \cite{Blandeau2011}. 
It is clear that an independent formulation is needed to explain these discrepancies and identify the correct form of Schlinker and Amiet's theory. 
 
This paper presents a new formulation of trailing edge noise for small rotating blade elements~(section~\ref{cha:kim-george-approach}). It is analogous to the formulations developed by~\citet{glegg1998broadband} for ducted fans, although cascade effects are not taken into account. The derivation starts with Goldstein's~(\citeyear{Goldstein1976}) expression for loading noise and uses the Green's function of \citet{garrick_theoretical_1954}, which incorporates the effect of mean flow, to derive an expression for pressure in the frequency domain. This expression is then used to derive the instantaneous PSD and the time averaged PSD. The new formulations generalize the compact source solution of~\cite{Kim1982}, who used the expression of~\cite{FfowcsWilliams1969} for a rotating point load. They incorporate an accurate description of the blade rotation, hence of the effect of blade acceleration; this effect is not taken into account in Schlinker and Amiet's formulation where the blade is assumed to be in uniform rectilinear motion. The time averaged formulation can therefore serve to explore the validity of Schlinker and Amiet's approach. 

The derivation of the new formulation was inspired by~\cite{Blandeau2011a} who first compared Schlinker and Amiet's model with a solution of the FWH equation. Their formulation of Schlinker and Amiet's model differed significantly from that of~\cite{Schlinker1981a} but was not accompanied by an alternative derivation. Also, they neglected the effect of mean flow and spanwise wavenumber. Both effects are taken into account in our new formulation. 

We revisit Schlinker and Amiet's theory and present simple derivations for some key steps that help to obtain the correct result (section \ref{sec:amiets-appr-rotat}). Of particular interest is the derivation showing how a theory derived for a stationary aerofoil can be generalized to an aerofoil in uniform motion. 
Finally, the results are validated by comparing the new formulation to Schlinker and Amiet's for multiple test cases, including a wind turbine, a cooling fan and an aircraft propeller (section~\ref{sec:results}).

\printnomenclature
\section{New formulation of trailing edge noise for rotating blades}
\label{cha:kim-george-approach}
\begin{figure}[htp]
  \centering
  \begin{minipage}{0.45\textwidth}
    \includegraphics[height=4.5cm]{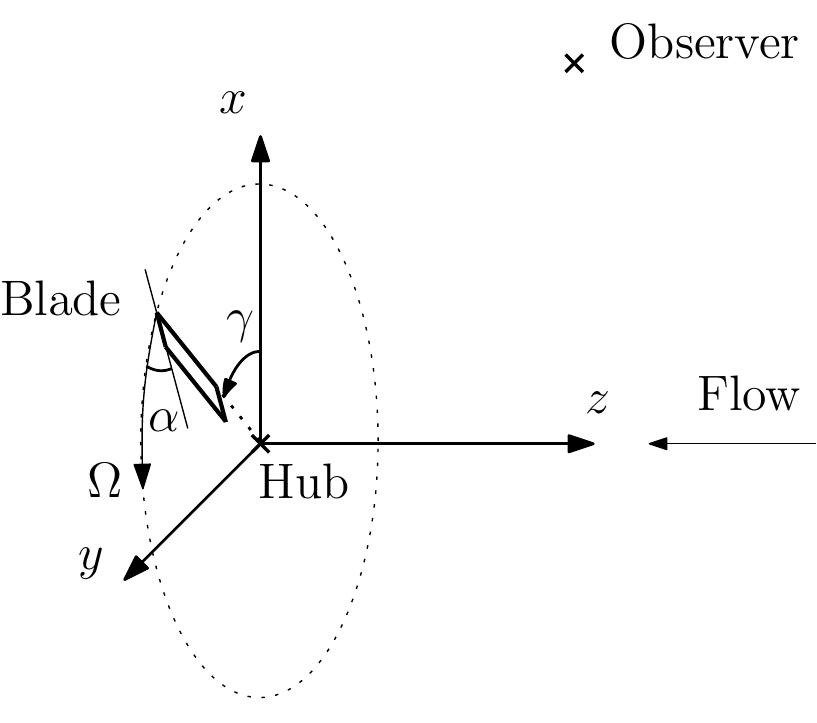}
    \caption{Stationary rotor and observer in a uniform flow. The reference
        frame is attached to the hub around which a blade element is
        rotating at angular velocity $\Omega$.}
    \label{fig:rotor_reference_frame}
  \end{minipage}
\quad%
  \begin{minipage}{0.45\textwidth}
    \includegraphics[height=4.5cm]{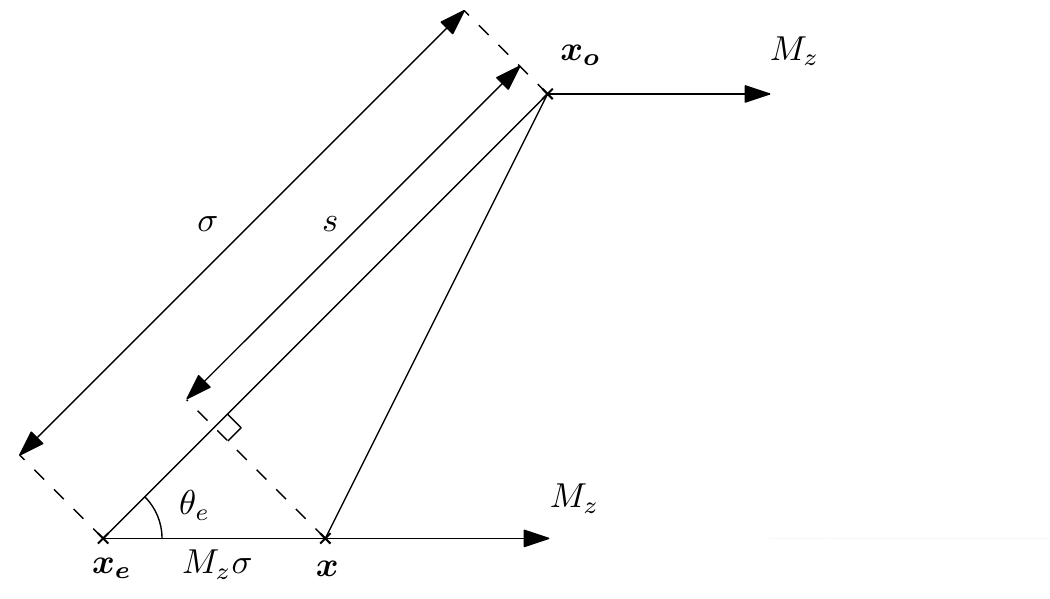}
    \caption{Fluid-fixed coordinates: source and observer ($\mbf{x_o}$) are moving at Mach $M_z$ in a quiescent medium. The source locations are $\mbf x$ at current time and $\mbf{x_e}$ at emission time.%
}
    \label{fig:garrick_watkins}
  \end{minipage}
\end{figure}
\subsection{Pressure field in the time domain}
Consider the problem of trailing edge noise radiation in a uniform flow from a stationary rotor towards a stationary observer. Although the results of this section will be applicable to propeller blades or fans, the rotor is described as a wind-turbine whereby the blade is being rotated by the incoming flow (figure~\ref{fig:rotor_reference_frame}). To derive the pressure field in the frequency domain we start by expressing the pressure field in the time domain. From Goldstein~(\citeyear{Goldstein1976}), loading noise can be expressed as 
\begin{equation}
  \label{eq:pressure_solution0}
  p(\mbf{x_o}, t) = -\intx{-T}{T}{\intx{\Sigma}{}{\gradient G(\mbf{x}, \tau | \mbf{x_o}, t) \cdot \mbf{L}(\mbf x, \tau) }{\Sigma}}{\tau},
\end{equation}
\nomenclature{$\mbf{x_o}$}{Observer position}%
\nomenclature{$G(\mbf x, t | y, \tau)$}{Time domain free field Green's function for source $(y, \tau)$ and observer $(\mbf x, t)$}%
\nomenclature{$\mbf L(\mbf x, t)$}{Unsteady lift force per unit area on a flat-plate aerofoil}%
where $-T < \tau < T$ is the interval of time over which sound is emitted ($T$ will be allowed to become infinitely large), 
$\Sigma$ is the blade planform and $G$ a free field Green's function that satisfies the convected wave equation
\nomenclature{$\gradient{_o}$}{Gradient relative to $\mbf{x_o}$}%
\nomenclature{$G(\mbf{x}, \tau | \mbf{x_o}, t)$}{Green's function for the convective wave equation in the time domain}%
\begin{equation}
  \label{eq:wavequation_time}
  \left(\gradient{_o^2} - \frac{1}{c_0^2} \Deli{_o^2}{t^2} \right) G(\mbf{x}, \tau | \mbf{x_o}, t) = - \delta(t - \tau) \delta(\mbf{x_o} - \mbf x),
\end{equation}
where $D_o/Dt = \del/\del t - U_z \del / \del z_o$ (with $U_z>0$ in the minus $z$-direction as shown in figure \ref{fig:rotor_reference_frame}). 
\nomenclature{$U_z$}{Flow velocity in the minus $z$-direction (figure 1)}%
Compared to alternative methods, a convected Green's function simplifies the derivation by taking care of sound waves convection by the flow. 
\cite{garrick_theoretical_1954} derived an expression for $G$ that is analogous to the free field Green's function,
\begin{equation}
  \label{eq:G_garrick_watkins}
  G(\mbf{x}, \tau | \mbf{x_o}, t) = \frac{\delta(t - \tau - \sigma / c_0)}{4\pi s},
\end{equation}
where $\mbf{x}$, $\mbf{x_o}$, $t$ and $\tau$ denote respectively the (present) source position, the observer position, the reception time and the emission time, and where the phase radius $\sigma$ and amplitude radius $s$ are defined as
\begin{align}
  \label{eq:garrick_watkins_notations}
  \sigma &= (M_z (z_0 - z) + s) / \beta^2, & s^2 = (z_0 - z)^2 + \beta^2 [(y_0 - y)^2 + (z_0 - z)^2],
\end{align}
\nomenclature{$s$}{Amplitude radius (equation \eqref{eq:garrick_watkins_notations})}%
\nomenclature{$\sigma$}{Phase radius (equation \eqref{eq:garrick_watkins_notations})}%
where $\beta=\sqrt{1 - M_z^2}$. The geometric interpretation of $\sigma$ and $s$ (\cite{garrick_theoretical_1954}) is given in figure~\ref{fig:garrick_watkins}, for the equivalent problem of a rotor and observer moving at Mach $M_z \mbf{\hat z}$ in a quiescent medium. 

The Green's function takes a simple form in the far~field (\cite{hanson_sound_1995}) when the source position is expressed in cylindrical coordinates $(r, \gamma, z)$ and the observer position in spherical emission coordinates $(R_e, \theta_e, \gamma_0)$, where $R_e$ corresponds to $\sigma$ in figure~\ref{fig:garrick_watkins} for $\mbf x = 0$ (present source at the hub): 
\begin{align}
  \label{eq:far_field_sigma_S}
  \sigma &\approx R_e - z_D \cos \theta_e - r_D \sin \theta_e \cos(\gamma - \gamma_0),  & s \approx R_e (1 - M_z \cos \theta_e),
\end{align}
\nomenclature{$(~)_D$}{Subscript denoting division by Doppler shift $1 - M_z \cos \theta_e$}%
where the $D$ subscript means that the variable has been divided by the Doppler factor $1 - M_z \cos \theta_e$. The above expressions can be obtained by applying Taylor's approximation to equation~\eqref{eq:garrick_watkins_notations} about the origin of the $|\mbf x|$ coordinate system, retaining terms up to order 0 for $s$ and 1 for $\sigma$, and by converting the observer position to emission coordinates.
\subsection{Pressure field in the frequency domain} 
Since Amiet's trailing edge noise theory is expressed in the frequency domain, we seek a general expression for pressure in the frequency domain. Taking the Fourier transform over observer time $t$ of~\eqref{eq:pressure_solution0} yields 
\begin{equation}
  \label{eq:pressure_solution_frequency}
  \tilde p(\mbf{x_o}, \omega) = -\intx{-T}{T}{\intx{\Sigma}{}{\mbf \gradient \tilde G(\mbf{x}, \tau | \mbf{x_o}, \omega) \cdot \mbf{L}(\mbf x, \tau) }{\Sigma}}{\tau},
\end{equation}
where the Fourier transform pair $(f, \tilde f)$ is defined as
\begin{align}
  \label{eq:FT_convention}
  \tilde f(\omega) &= \frac{1}{2\pi}  \intx{-\infty}{\infty}{f(t) e^{-i \omega t}}{t}, & f(t) &= \intx{-\infty}{\infty}{\tilde f(\omega) e^{i \omega t}}{\omega}.
\end{align}

The frequency domain Green's function $\tilde G$ is obtained by taking the Fourier transform over $t$ of~\eqref{eq:G_garrick_watkins} and incorporating the far field approximation~\eqref{eq:far_field_sigma_S}. The result is further decomposed into a series of azimuthal modes~(see appendix~\ref{sec:far-field-greens}) to simplify the computation of the gradient, 
\begin{align}
  \label{eq:Gw}
  \tilde G (\mbf x, \tau | \mbf{x_o}, \omega) &= \tilde g(\omega/c_0, R_e, \theta_e) \sum_{n=-\infty}^{+\infty}  J_n(k_r r) e^{-i n (\gamma_0 - \pi/2)} e^{-i(\omega \tau - n \gamma - k_z z)}, \\
  \label{eq:g}
  \tilde g(k, R_e, \theta_e) &= \frac{e^{ -i k R_e}}{8\pi^2 R_e(1-M_z \cos \theta_e)}, 
\end{align}
\nomenclature{$k_r, k_z$}{Doppler shifted acoustic wavenumbers defined below equation (2.9).}%
where we have introduced the axial wavenumber $k_z = k_D \cos \theta_e$ and the radial wavenumber $k_r = k_D \sin \theta_e$.  

\begin{figure}[hbt]
  \centering
  \subfigure[3D view]{\includegraphics[width=.6\textwidth]{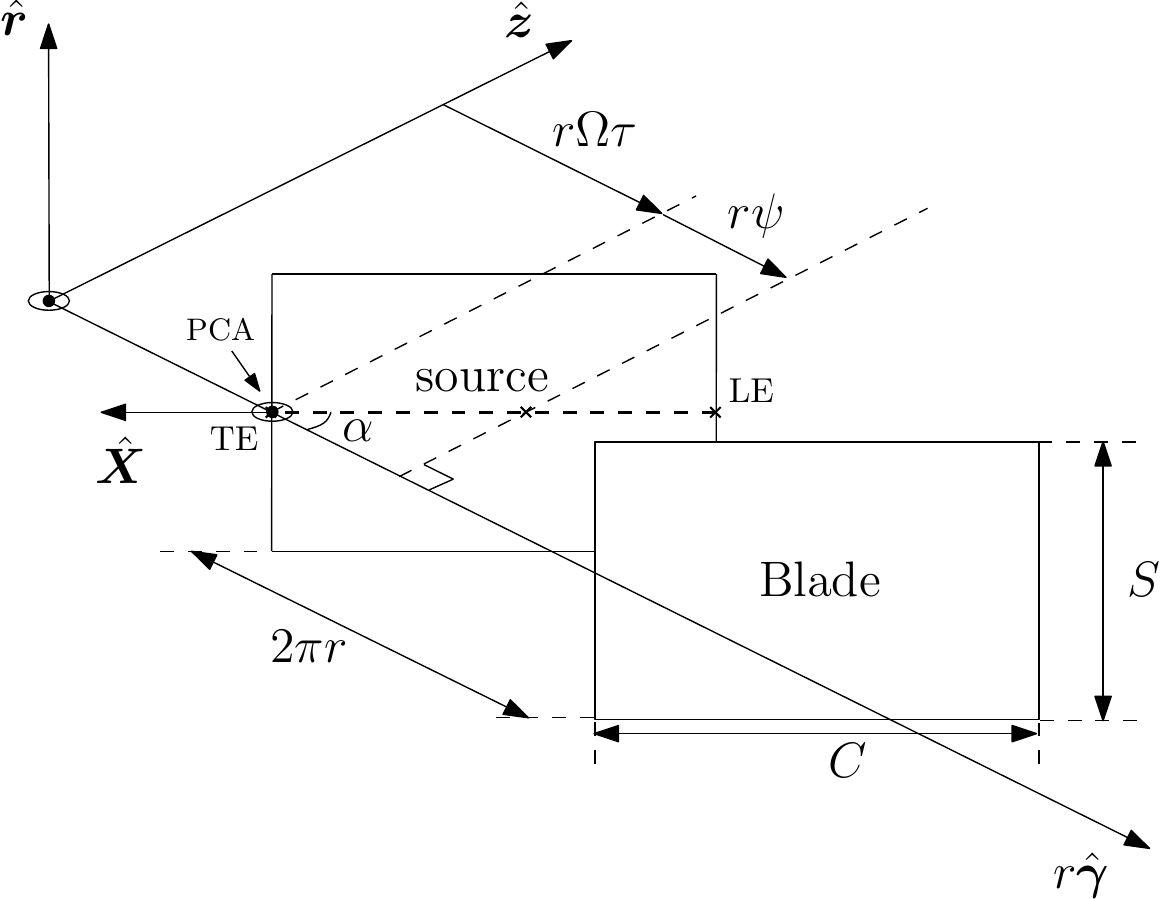}}\quad
  \subfigure[2D view of blade section]{\includegraphics[width=.35\textwidth]{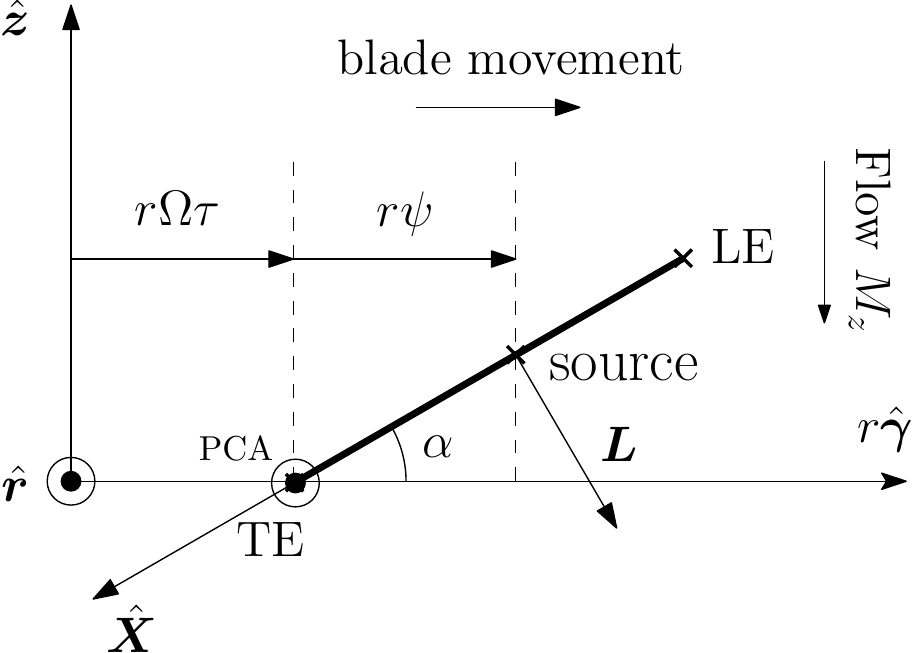}}
  \caption{Blade geometry and source position. The blade path is unwrapped in the azimuthal direction $\hat{\mbf{\gamma}}$: it moves at speed $r\Omega$ along the $\hat{\mbf\gamma}$-axis. The pitch change axis (PCA) is located at the trailing edge (TE) and indicates the axis around which the blade surface is rotated through pitch angle $\alpha$ relative to the rotor plane $(\hat{\mbf{r}}, \hat{\boldsymbol{\gamma}})$. The source position is $(r, \Omega \tau + \psi, z)$ in the hub-fixed cylindrical coordinate system (figure~\ref{fig:rotor_reference_frame}) and $(r, X)$ in the blade-fixed coordinate system.}
  \label{fig:blade_coordinates}
\end{figure}

From figure~\ref{fig:blade_coordinates}(b), the lift lies in the plane ($\mbf{\hat \gamma}, \mbf{\hat z}$) and can be expressed in terms of the magnitude $L$ and the pitch angle $\alpha$. The dot product $\scalar{L}{\gradient \tilde G}$ can be expressed in terms of the gradients of $\tilde G$ in the azimuthal and axial directions that are easily derived from~\eqref{eq:Gw}. Substituting the expression for the dot product into~\eqref{eq:pressure_solution_frequency} yields a harmonic pressure of the form
\begin{multline}
  \label{eq:pressure_solution1}
\tilde p(\mbf{x_o}, \omega) =  \intx{\Sigma}{} {i \tilde g
    \sum_{n=-\infty}^{+\infty}  k_L J_n e^{i [k_z z - n (\gamma_0 - \pi/2)] }\intx{-T}{T}{L e^{-i(\omega \tau - n \gamma)}} {\tau}}{\Sigma},
\end{multline}
where $k_L = k_{z} \cos \alpha - (n/r) \sin \alpha$ is the magnitude of a modal wavenumber orthogonal to the blade planform.
\nomenclature{$k_L, k_C$}{Chordnormal (equation~\eqref{eq:kL_kC_phi}) and chordwise (equation~\eqref{eq:kL_kC_phi}) acoustic wavenumber respectively}%

We now turn to the rotational motion of the source. We will assume that the pitch change axis (PCA) of the blade is at the trailing edge (TE) and that the trailing edge alignment and face alignment (\cite{hanson_influence_1980}) are 0. In other words, lean and sweep are both neglected. The geometry of the blade is illustrated in figure~\ref{fig:blade_coordinates}. From figure~\ref{fig:blade_coordinates}(b) the azimuthal angle of a source within the blade planform is $\gamma = \Omega \tau + \psi$. If $T$ tends to infinity, the time integral in~\eqref{eq:pressure_solution1} reduces to $2\pi \tilde L(\bm{x}, \omega - n\Omega) e^{in\psi}$: the effect of the blade rotation is to select, for each azimuthal mode, the frequency $\omega - n\Omega$ from the spectrum of the blade loading, so
\begin{align}
  \label{eq:pressure_solution2_0}
\tilde p(\mbf{x_o}, \omega) &=   2\pi i \tilde g
    \sum_{n=-\infty}^{+\infty}  e^{-i n (\gamma_0 - \pi/2)}\intx{\Sigma}{}{k_L J_n  \tilde L(\mbf x, \omega - n\Omega) e^{i (k_z z + n \psi ) } }{\Sigma}.
\end{align}

Finally, the lift $\tilde L(\mbf x, \omega)$ is given by Amiet's theory (detailed in section \ref{sec:trailing-edge-noise}): for every spanwise wavenumber $K_r$, a gust of amplitude $P_0(\omega, K_r)$ travelling at convection speed $U_c$ and scattered at the trailing edge, gives rise to a pressure jump $P_0(\omega, K_r) g_0(X, \omega/U_c, K_r)$ along the chord, where $g_0$ is the blade response function (Amiet \citeyear{Amiet1976}, \citeyear{Amiet1978}) and $X$ is the chordwise source position measured from the trailing edge ($-C \leq X \leq 0$ as shown in figure~\ref{fig:blade_coordinates}). Following Amiet, the gust amplitude $P_0$ is a function of angular frequency $\omega$ and radial wavenumber $K_r$. Here, we express the amplitude as a function $P(K_X, K_r)$ of chordwise wavenumber and radial wavenumber. Since the chordwise wavenumber of the gust equals $\omega/U_c$, it is easy to show that $P(\omega/U_c, K_r) = U_c P_0(\omega, K_r)$. The blade loading due to trailing edge noise is therefore given by
\nomenclature{$P(K_X, K_r)$}{Wall pressure in the wavenumber domain}%
\nomenclature{$K_X, K_r$}{Chordwise and spanwise hydrodynamic wavenumbers}%
\begin{equation}
  \label{eq:lift_2}
  \tilde L(\mbf x, \omega) = \intxInf{\frac{1}{U_c} P(\omega / U_c, K_r) g_0(X, \omega/U_c, K_r)  e^{-i K_r r}}{K_r},
\end{equation}
\nomenclature{$X$}{Source position along the chord from the trailing edge ($X \leq 0$)}%

As illustrated in figure~\ref{fig:blade_coordinates} , the source coordinates $(\psi, z)$ can be expressed in terms of $X$ as 
\begin{align}
  \label{eq:blade_coordinates}
  r \psi &= - X \cos \alpha, & z &= - X \sin \alpha.
\end{align}
\nomenclature{$\psi$}{Source azimuthal angle relative to the trailing edge}%
The phase $k_z z + n \psi = -k_C X$, where $k_C$ is a chordwise acoustic wavenumber:
\begin{equation}
  \label{eq:kL_kC_phi}
\column{k_L}{k_C} = \mymatrix{\cos \alpha}{-\sin \alpha}{\sin \alpha}{\cos \alpha} \column{k_z}{n/r}.
\end{equation}
Note how the acoustic wavenumbers in blade coordinates $(k_L, k_C)$ are obtained by rotating ${k_z \mbf{\hat z} + (n/r) \mbf{\hat \gamma} }$ by angle $\alpha$ around the PCA. After expressing the lift in terms of Amiet's blade response function using~\eqref{eq:lift_2} and since $d\Sigma = dX dr$ in \eqref{eq:pressure_solution2_0}, we find that the impact of the lift on the pressure field is measured through the acoustically weighted lift $\Psi_L$ defined as
\begin{equation}
  \label{eq:acoustic_lift_from_g0}
  \Psi_L(K_X, K_r, k_C) = \frac{2}{C}\intx{-C}{0}{g_0(X, K_X, K_r) e^{-i k_C X}}{X},
\end{equation}
for which an analytical expression will be provided in equation~\eqref{eq:Psi_L_expression}. The frequency domain pressure at the observer location $\mbf {x_o}$ can be expressed as
\begin{multline}
  \label{eq:pressure_solution1b}
\tilde p(\mbf{x_o}, \omega) =  \frac{  i \exp (-ik R_e)}{8\pi R_e (1 - M_z \cos \theta_e)} 
\sum_{n=-\infty}^{+\infty} \intx{r}{}{ 
  \intxInf{
\frac{k_L C}{U_c} J_n(k_r r) e^{-in (\gamma_0 - \pi/2)}\\
  \times  P(K_X, K_r) \Psi_L(K_X, K_r, k_C) e^{-i K_r r} }{K_r} 
}{r},
\end{multline}
where $K_X = (\omega - n \Omega)/U_c$.
\subsection{Instantaneous spectrum}
We now seek an expression for the instantaneous (Wigner-Ville) spectrum~\citep{flandrin_time-frequency/time-scale_1998} for direct comparison with Schlinker and Amiet's theory (equation~\eqref{eq:instant_Spp}). An autocorrelation function for the pressure $p$ at the observer position $\mbf{x_o}$ can be defined as
\begin{equation}
  \label{eq:autocorrelation}
  R_{pp}(\mbf{x_o}, t, \tau) = E[p(\mbf{x_o}, t + \tau/2) p^\star(\mbf{x_o}, t - \tau/2)],
\end{equation}
where $E$ denotes the expected value and $\star$ the complex conjugate. Since the blade element is rotating around the axis, this function is periodic in $t$ with period $T_\Omega$.
The instantaneous spectrum $S_{pp}(\mbf{x_o}, \omega, t)$ is defined as the Fourier transform of $R_{pp}(\mbf{x_o}, t, \tau)$ over $\tau$. An equivalent definition is
\begin{equation}
  \label{eq:instant_spectrum_kg}
  S_{pp}(\mbf{x_o}, \omega, t) = \intxInf{\tilde R_{pp}(\mbf{x_o}, \omega, \varpi) e^{i\varpi t}}{\varpi},
\end{equation}
where $\tilde R_{pp}(\mbf{x_o}, \omega, \varpi) \equiv E[\tilde p^\star(\mbf{x_o}, \omega - \varpi/2) \tilde p(\mbf{x_o}, \omega + \varpi/2)]$ is a cross-correlation in the spectral domain. Equation~\eqref{eq:instant_spectrum_kg} states that the instantaneous spectrum is the inverse Fourier transform over frequency $\varpi$ of $\tilde R_{pp}(\mbf{x_o}, \omega, \varpi)$.
\subsubsection{Spectral cross-correlation $\tilde R_{pp}(\mbf x_o, \omega, \varpi)$}
From~\eqref{eq:pressure_solution1b}, the spectral cross-correaltion is given by
\begin{multline}
  \label{eq:Spp2}
  \tilde R_{pp}(\mbf{x_o}, \omega, \varpi) = \frac{e^{-i (\varpi/c_0) R_e}}{\left[ 8 \pi R_e (1 - M_z \cos \theta_e)\right]^2}  %
\sum_{n=-\infty}^{+\infty} \sum_{n'=-\infty}^{+\infty} e^{i (n - n') (\gamma_o - \pi/2)} \times \\%
\intx{R_h}{R_t}{
  \intx{R_h}{R_t}{
    \intxInf{
      \intxInf{
        \frac{k_L C}{U_c} \frac{k_L' C'}{U_c'} e^{i(K_r r - K_r' r')} \Psi_L^* \Psi_L' J_n( k_r r) J_{n'}(k_r' r') \\
        E[P^\star(K_X, K_r) P(K_X', K_r')]
      }{K_r'}
    }{K_r}
  }{r'}
}{r},
\end{multline}
where the primed wavenumbers and non-primed wavenumbers are associated respectively with frequency $\omega + \varpi/2$ and $\omega - \varpi/2$.

We will now derive two approximate expressions $\tilde R_{pp}^{(1)}$ and $\tilde R_{pp}^{(2)}$ for $R_{pp}$, by simplifying the integrals over $r'$ and $r$ respectively. These two expressions will then be combined to give an expression for $R_{pp}$ that has Hermitian symmetry; this is necessary for the instantaneous PSD to be real. 

From~\cite{Blandeau2011} (appendix D), if we assume that the flow is approximately 2D and that it is uniform across the span (strip theory assumption) then the two gusts are correlated only when their wavenumbers are equal: $K_r = K_r'$ and $K_X = K_X'$ (i.e., from the definition of $K_X$ below equation~\eqref{eq:pressure_solution1b}, if $\varpi = (n' - n)\Omega$), and we have 
\begin{equation}
  \label{eq:surface_psd}
  E[P^\star(K_X, K_r) P(K_X', K_r')] = U_c' \, \delta(K_r' - K_r) \, \delta\left(\varpi - (n' - n)\Omega\right)  \Phi_{qq}(K_X', K_r'),
\end{equation}
\nomenclature{$\delta_{m, n}$}{Kronecker delta: 1 if $m=n$, 0 otherwise}%
where $\Phi_{qq}$ is the wavenumber spectrum of the aerofoil surface pressure produced by the turbulence. The above equation assumes that the convection velocity is constant over the area where $P(K_X, k_r)$ and $P(K_X', k_r')$ are correlated. For a given radius $r$, that region is defined by $r - \Delta r \leq r' \leq r + \Delta r$, where $\Delta r$ is slightly larger than the turbulence correlation length $l_S$. Since $U_c' = r' \Omega$, variations in convection velocity are negligible over that region provided that $\Delta r \ll r'$, i.e. $l_S \ll r'$. Using Corco's equation~\eqref{eq:ly}, at high frequency, $l_S \approx 0.8 r' \Omega / (\omega' \eta)$ (where $\eta$ is a constant of order one) so the necessary condition becomes $\Omega \ll \omega'$: the rotation speed should be much smaller than the source frequency. This corresponds to the assumption made by Amiet (see introduction of section~\ref{sec:amiets-appr-rotat}).
\nomenclature{$\Phi_{qq}(k_X, k_S)$}{Surface pressure spectrum close to the trailing edge}%

Substituting \eqref{eq:surface_psd} in~\eqref{eq:Spp2} leads to $\tilde R_{pp} = \tilde R_{pp}^{(1)}$, where
\begin{multline}
  \label{eq:Rpp}
  \tilde R_{pp}^{(1)}(\mbf{x_o}, \omega, \varpi) \equiv B(\varpi)  %
  \intx{R_h}{R_t}{%
      \sum_{n=-\infty}^{+\infty}  
      \sum_{n'=-\infty}^{+\infty} e^{i(n-n')(\gamma_o - \pi/2)} I_{n, n'}' \times  \\  
\delta\left(\varpi - (n' - n)\Omega\right)}{r},
\end{multline}
\begin{align}
\label{eq:amplitude}
&B(\varpi) = \frac{e^{-i \varpi R_e / c_0}}{\left[ 8 \pi R_e (1 - M_z \cos \theta_e)\right]^2},\\
\label{eq:In_init}
&I_{n, n'}'(\omega, \varpi, r) \equiv \intxInf{L_{n, n'}(K_r') \Phi_{qq}(K_X', K_r')}{K_r'}, \\
\label{eq:In1}
  &L_{n,n'}(K_r') \equiv \intx{R_h}{R_t}{
      \frac{k_L C k_L' C'}{U_c} \Psi_L^* \Psi_L'
      J_n\left(k_r r\right) J_{n'}\left(k_r' r' \right) e^{i K_r'(r - r')}
    }{r'}.
\end{align}
The integration over $r'$ may be truncated at $r' = r \pm \Delta r$. 
In this narrow strip, the flow variables can be considered constant in the amplitude terms of~\eqref{eq:In1} ($k_L$, $k_L'$, $U_c$, $U_c'$, $C$ and $C'$): they are equal to their values at $r'=r$.
We seek to simplify~\eqref{eq:In1} and put it in an exponential form to capture its phase. This is needed to analyse the spanwise behaviour of the solution. The main issue is with $\Psi_L$, which may be studied using a power series expansion. However, for simplicity, we will neglect the phase changes in $\Psi_L$ and assume $\Psi_L'(r') \approx \Psi_L'(r)$ so that $L_{n,n'}$ takes the form
\begin{equation}
  \label{eq:I}
  L_{n,n'}(K_r') \approx \frac{k_L k_L'C^2}{U_c} \Psi_L^\star(K_X, K_r', k_C) \Psi_L(K_X', K_r',  k_C') J_n\left(K_r' r\right) P_{n,n'}(K_r'),
\end{equation}
where, making the change of variable $\eta=r'-r$,
\begin{equation}
  \label{eq:Pm}
  P_{n,n'}(K_r') \approx \intx{-\Delta r}{\Delta r}{%
      J_{n'}\left(k_r' (\eta + r) \right) e^{-iK_r' \eta}
  }{\eta}.%
\end{equation}
The Bessel function in~\eqref{eq:Pm} can be rewritten as an integral~(\cite{abramowitz_bessel}):
\begin{equation}
  \label{eq:Jm_integral}
  J_{n'}(k_r'(\eta + r)) \equiv \frac{1}{2\pi} \intx{-\pi}{\pi}{e^{i (n' \gamma - k_r' (\eta + r) \sin \gamma)}}{\gamma}.
\end{equation}
Substituting \eqref{eq:Jm_integral} into \eqref{eq:Pm},
\begin{equation}
  \label{eq:Pm2}
  P_{n,n'}(K_r') = \frac{1}{2\pi} \intx{-\pi}{\pi}{
    e^{i (n' \gamma - k_r' r \sin \gamma)} \intx{- \Delta r}{+ \Delta r }{
      e^{-i (K_r' - K_{r0}') \eta}
    }{\eta}
  }{\gamma},
\end{equation}
where $K_{r0}' = -k_r' \sin \gamma$.
\nomenclature{$K_{r0}$}{Defined below equation (2.28)}%
If $\Delta r$ is large relative to the wavelengths of interest in this
problem, i.e. if ${K_r'\Delta r \gg 1}$, then the second integral in
\eqref{eq:Pm2} simplifies to
\begin{equation}
  \label{eq:delta_simplification}
  \intx{- \infty}{+\infty }{e^{-i (K_r' - K_{r0}') \eta}}{\eta} = 2\pi \delta(K_r' - K_{r0}'),
\end{equation}
The above simplification may not always be possible in which case the integral can be solved numerically. 
Substituting~\eqref{eq:delta_simplification} into~\eqref{eq:Pm2} to get $P_{n,n'}(K_r')$, and combining the result with~\eqref{eq:I}~and~\eqref{eq:In_init} yields
\begin{equation}
\label{eq:In_old}
I_{n, n'}'(\omega, \varpi, r) \approx \frac{k_L k_L'C^2}{U_c} J_n\left(k_r r\right) Q_{n',n}'(\omega, \varpi, r),
\end{equation} 
where
\begin{equation}
  \label{eq:Q'}
Q_{n',n}' = \intx{-\pi}{\pi} { 
            e^{i (n' \gamma - k_r' r \sin \gamma)}
            \Psi_L^\star(K_X,  K_{r0}', k_C) \Psi_L(K_X', K_{r0}', k_C') \Phi_{qq}(K_X', K_{r0}')
          }{\gamma}.
\end{equation}

We can derive a different expression $\tilde R_{pp}^{(2)}$ for $\tilde R_{pp}$, by approximating the integral over $r$ in equation~(\ref{eq:Spp2}) and by following the steps in equations (\ref{eq:Rpp}) to (\ref{eq:Q'}) to get
\begin{multline}
  \label{eq:Rpp2}
  \tilde R_{pp}^{(2)}(\mbf{x_o}, \omega, \varpi) \equiv B(\varpi)  %
  \intx{R_h}{R_t}{%
      \sum_{n=-\infty}^{+\infty}  
      \sum_{n'=-\infty}^{+\infty} e^{i(n-n')(\gamma_o - \pi/2)} I_{n, n'} \times  \\  
\delta\left(\varpi - (n' - n)\Omega\right)}{r},
\end{multline}
where
\begin{align}
\label{eq:In'n}
  &I_{n', n}(\omega, \varpi, r') \approx \frac{k_L k_L'C^2}{U_c} J_{n'}\left(k_r' r'\right) Q_{n, n'}(\omega, \varpi, r'), \\
\label{eq:Q}
  &Q_{n, n'} = \intx{-\pi}{\pi} { 
            e^{-i (n\gamma - k_r r' \sin \gamma)}
            \Psi_L^\star(K_X,  K_{r0}, k_C) \Psi_L(K_X', K_{r0}, k_C') \Phi_{qq}(K_X, K_{r0})
          }{\gamma},
\end{align}
where $K_{r0}=-k_r \sin \gamma$ and the wavenumbers $k_L$, $k_L'$, $k_C$ and $k_C'$ are evaluated at radius $r'$. Note that in deriving $K_{r0}$ we have used the conjugate form of equation~(\ref{eq:Jm_integral}) to express $J_{n}(k_rr)$. We have also chosen this time $K_X$ instead of $K_X'$ as the first argument of $\Phi_{qq}$. This is justified since $K_X$ equals $K_X'$.

If we change $\varpi$ to $-\varpi$ and exchange $n$ and $n'$ then $Q'_{n',n}$ turns into $Q_{n,n'}^{\star}$. This is sufficient to show that $R_{pp}^{(1)}(\mbf{x_o}, \omega, -\varpi, r)$ equals the conjugate of $R_{pp}^{(2)}(\mbf{x_o}, \omega, \varpi)$. We express $\tilde R_{pp}$ as 
\begin{equation}
  \label{eq:RppHermitian}
  \tilde R_{pp} \approx \frac{\tilde R_{pp}^{(1)} + \tilde R_{pp}^{(2)}}{2},
\end{equation}
which satisfies the Hermitian symmetry required for the instantaneous spectrum to be real valued.

\subsubsection{Instantaneous spectrum $S_{pp}(\mbf x_o, \omega, t)$}
Substituting equations~(\ref{eq:Rpp}) and (\ref{eq:Rpp2}) into (\ref{eq:RppHermitian}) and then into~\eqref{eq:instant_spectrum_kg},  
\begin{multline}
  \label{eq:instant_spectrum_kg_final}
  S_{pp}(\mbf{x_o}, \omega, t) = 
      \intx{R_h}{R_t}{
          \sum_{m=-\infty}^{+\infty}  B(m\Omega) e^{-i m(\gamma_o - \pi/2)} \times \\
          \left(\sum_{n=-\infty}^{+\infty}  
           \tilde I_{n, n + m}(\omega, m\Omega, r) \right)  e^{i m \Omega t}  }{r},
\end{multline}
where we have made the change of variable $m = n' - n$ and ${\tilde I_{n,n+m} \equiv (I_{n,n+m}' + I_{n+m,n})/2}$.

Note that the expression of equation~(\ref{eq:instant_spectrum_kg_final}) is always real but it can become negative due to a known defect of the Wigner-Ville spectrum, which does not define a true spectral density. 
\subsubsection{Time averaged PSD $\overline S_{pp}(\mbf x_o, \omega)$}
The time averaged PSD is obtained by retaining the mode $m=0$ in~\eqref{eq:instant_spectrum_kg_final}, so $\varpi=0$ and the prime and non-prime quantities are at the same frequency $\omega$, and since $I_{n,n}^{\star}=I_{n,n}'$, we can write
\begin{equation}
  \label{eq:time_averaged_spectrum_kg}
  \overline S_{pp}(\mbf{x_o}, \omega) = 
  \intx{R_h}{R_t}{ B(0) \sum_{n=-\infty}^{+\infty} \tilde I_{n}(\omega, r) }{r},
\end{equation}
where, from~(\ref{eq:In_old}) and~\eqref{eq:In'n},
\begin{multline}
\label{eq:In}
\tilde I_{n}(\omega, r) = \frac{k_L^2}{U_c}C^2 J_n\left(k_r r\right) 
           \intx{-\pi}{\pi} { 
            \cos{(n \gamma - k_r r \sin \gamma)} \times \\
            |\Psi(K_X, -k_r \sin \gamma, k_C)|^2 
            \Phi_{qq}(K_X, -k_r \sin \gamma)}{\gamma}. 
\end{multline}
Our numerical experiments using the above equations indicate that $\overline S_{pp}$ is always positive. 
 
For a small blade element of span $S$, 
\begin{equation}
  \label{eq:Spp_final_simple}
  \overline S_{pp} (\mbf{x_o}, \omega) = %
          B(0) S \sum_{n=-\infty}^{+\infty}  \tilde I_{n}(\omega, r).
\end{equation}
\section{Trailing edge noise theory for isolated aerofoils}
\label{sec:trailing-edge-noise}
\begin{figure}[hbp]
  \centering
  \includegraphics[]{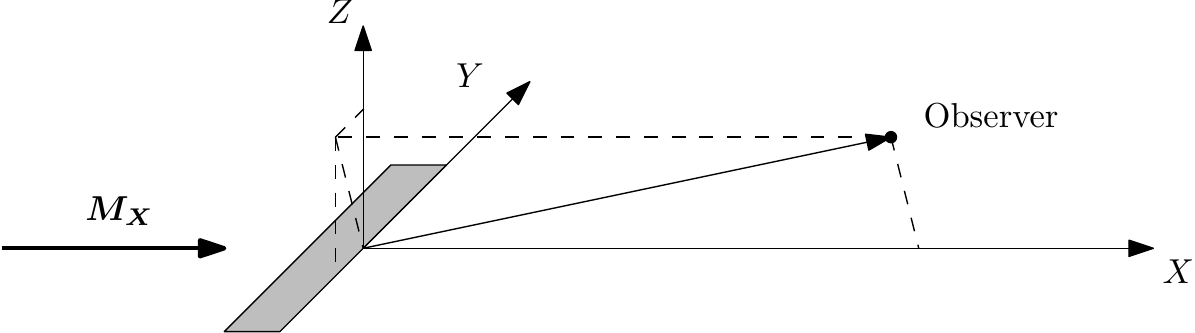}
  \caption{Coordinate system for an isolated aerofoil in a wind tunnel. The aerofoil and observer are stationary. The flow is uniform at zero angle of attack and Mach $\mbf{M_X}$.}
  \label{fig:isolated_aerofoil_frame}
\end{figure}

\nomenclature{$\omega$}{Angular frequency of the sound at the observer location}%
\nomenclature{$S_{pp}(\omega)$}{Pressure PSD at the observer location}%
\nomenclature{$c_0$}{Speed of sound}%
\nomenclature{$C$}{chord}%
\nomenclature{$S$}{span}%
\nomenclature{$(X, Y, Z)$}{Blade coordinate system (see fig. \ref{fig:isolated_aerofoil_frame} and \ref{fig:coordinates_isolated})}%
\nomenclature{$k_S$}{Spanwise acoustic wavenumber (equation~(\ref{eq:amiet_psi_params})}%
\nomenclature{$\Psi(k_X, k_S, k_C)$}{Acoustically weighted lift}%
\nomenclature{$l_S$}{Spanwise correlation length}%
Consider a flat plate in a uniform flow of Mach number $M_X$ at zero angle of attack~(figure~\ref{fig:isolated_aerofoil_frame}).
The observer location is expressed using a cartesian coordinate system $(X, Y, Z)$, with $X$ in the chordwise direction and pointing downstream and $Z$ in the vertical direction.
From \cite{Amiet1976,Amiet1978}, using the large span assumption (equation~(\ref{eq:delta_simplification})), the PSD at frequency $\omega$ is given by
\begin{equation}
  \label{eq:psd_isolated}
  S_{pp}(\omega) = \left( \frac{\omega}{c_0} \frac{C}{2} \frac{Z}{2\pi s^2}\right)^2 \frac{S}{2}\frac{\pi}{U_c} \Phi_{qq}(k_X, k_S) |\Psi_L(k_X, k_S, k_C)|^2,
\end{equation}
where
\begin{align}
  \label{eq:psd_isolated_terms}
   s^2 &= X^2 + \beta^2 (Y^2 + Z^2), & \beta^2 &= 1 - M_X^2, & U_c &= 0.8 M_X c_0,
\end{align}
and from~\cite{Roger2005} the wavenumber spectrum and spanwise correlation length are given by
\begin{equation}
  \label{eq:Phiqq}
  \frac{\pi}{U_c}\Phi_{qq}(k_X, k_S) \equiv l_S(k_X, k_S) S_{qq}(\omega), 
\end{equation}
\nomenclature{$\mbf{M_X}$}{Mach number of the air relative to the blade at 0$^{\circ}$ angle of attack}%
\nomenclature{$U_c$}{Convection velocity of an eddy in the boundary layer near the trailing edge}%
\begin{align}
  \label{eq:ly}
  l_S(k_X, k_S) \equiv \frac{1}{k_X} \frac{\eta}{\eta^2 + (k_S / k_X)^2}.
\end{align}
\nomenclature{$\eta$}{Exponential decay rate of the spanwise coherence function}%
The term $\eta$ represents the exponential decay rate of the spanwise coherence function; \cite{Brooks1981a} measured $\eta=0.62$ for a NACA 0012 at Mach 0.11 and zero angle of attack. For simplicity, this value will be used for all Mach numbers in section~\ref{sec:results}.
For a stationary flat plate in a uniform flow (\cite{Amiet1975})
\begin{align}
  \label{eq:amiet_psi_params}
  k_X &= \frac{\omega}{U_c}, & k_S &= \frac{\omega}{c_0} \frac{Y}{s}, &k_C &= \frac{\omega}{c_{0} \beta^{2}} \left(M_X - \frac{X}{s}\right).
\end{align}

\subsection{Acoustic lift}					  
The acoustic lift, defined in equation~\eqref{eq:acoustic_lift_from_g0}, can be expressed as 
\begin{equation}
  \label{eq:Psi_L_expression}
  \Psi_L(k_X, k_S, k_C) = \frac{i}{A} \left\{ \frac{\sqrt{iB}}{\sqrt{iB - iA}} \erf{ \sqrt{2(iB - iA)}} + e^{i2A}\left[1 - \erf{\sqrt{2i B}}\right]\right\},
\end{equation}
\begin{align}
  \label{eq:A_B}
  A &= \overline k_X + \overline k_C, & B &= \overline k_X + \overline \kappa + M_X \overline \mu, & \mu = M_c k_X/ \beta^2,
\end{align}
where the overbar denotes normalisation by $C/2$ and $M_c = U_c / c_0$ the Mach number of the turbulent eddies in the boundary layer close to the trailing edge. The square roots in equation~(\ref{eq:Psi_L_expression}) have a branch cut along the negative real axis.

The wavenumber $\kappa$ is a function of the spanwise wavenumber $k_S$.
It is defined as 
\begin{equation}
\kappa \equiv \left\{ 
\begin{aligned}
\mu \sqrt{1 - k_S^{2} /(\beta \mu)^2} \quad &\text{if} &  k_S^2 < &(\beta \mu)^2 \\
-i |\mu| \sqrt{k_S^{2} /(\beta \mu)^2 - 1} \quad &\text{if} &  k_S^2 \geq &(\beta \mu)^2 \\
\end{aligned}\label{eq:5}
\right.
\end{equation}
so that the imaginary part of $\kappa$ is always negative.
This is required for the error functions in \eqref{eq:Psi_L_expression} to converge in the far field. The square roots in equations~\eqref{eq:5} and~\eqref{eq:Psi_L_expression} are classicaly defined with a branch cut along the negative real axis. 

\begin{figure}[hbt]
  \centering
  \includegraphics[height=5cm]{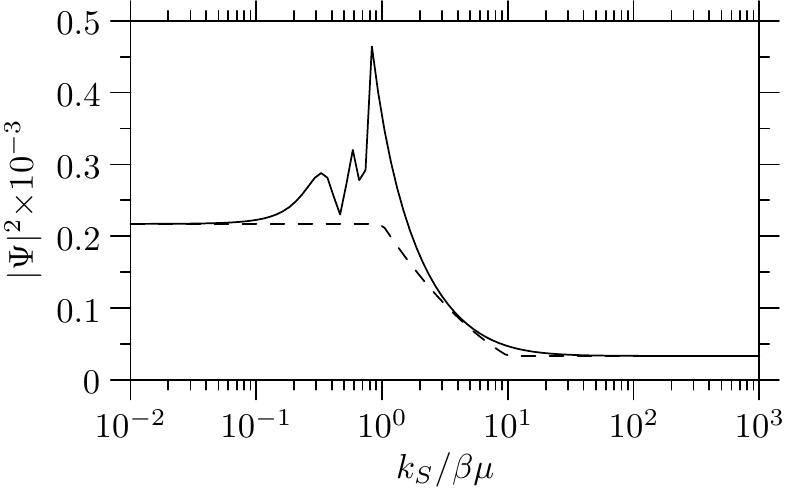}
  \caption{Acoustic lift spectrum $|\Psi(k_X, k_S, k_C)|^2$ as a function of spanwise wavenumber $k_S$, at Mach $M_X=0.1$ and frequency $f=1000\,\text{Hz}$.
The radiating wavenumber $k_C$ is defined assuming an observer at $x/s=0.7$.
The transition between supercritical and subcritical gusts occurs at $\xi \equiv k_S / (\beta \mu) = 1$.
The acoustic lift based on equation~\eqref{eq:Psi_L_expression} becomes unphysical near the supercritical-subcritical transition~(solid line, see~\cite{Roger2005}), so we assume that $|\Psi|$ is constant for $\xi < 1$ and $\xi > 10$, and use a linear interpolation as a function of $\log_{10}(\xi)$ for $1 \leq \xi \leq 10$ (dashed line).
}
  \label{fig:acoustic_lift_ky}
\end{figure}

Figure~\ref{fig:acoustic_lift_ky}, shows how $|\Psi|$ varies with spanwise wavenumber, for a constant frequency of 1 kHz (solid line).
The transition between supercritical and subcritical gusts occurs at $k_S / \beta \mu = 1$.
It can be seen that a sharp increase occurs near $k_S / \beta \mu = 1$.
That increase is non-physical and occurs because the governing equation reduces from a Helmholtz equation to a Laplace equation.
This was first pointed out by~\cite{Roger2005}.
Since $\Psi$ is asymptotically constant at low and high values of $k_S$ we define $\Psi$ as a piecewise function, such that it is constant for supercritical gusts ($k_S / \beta \mu < 1$), and for highly subcritical gusts ($k_S / \beta \mu > 10$).
Between those values, we use a linear interpolation of $|\Psi|$ in terms of $\log_{10} k_S / \beta \mu$.
The piecewise implementation of $|\Psi|^2$ is shown as a dashed line in the figure~\ref{fig:acoustic_lift_ky}.

\subsection{Surface pressure power spectral density}					  
\label{sec:surf-press-power}
The PSD of the incoming pressure fluctuations at the trailing edge, denoted $S_{qq}(\omega)$, can be measured experimentally.
If no experimental data is available, it can be estimated by using empirical low-order models.
These low order models are expressed in terms of parameters characterising the boundary layer, such as the boundary layer thickness, displacement thickness, wall shear stress etc.
A review of the different models is given by~\cite{Blandeau2011}. This paper uses the model of~\cite{Chou1984} which gives
\begin{align}
  \label{eq:Sqq}
  S_{qq}(\omega) = \left(\frac{1}{2} \rho U_X^2\right)^2  \frac{\delta^*}{U_X} F(\overline{\omega}), 
\end{align}
\nomenclature{$S_{qq}(\omega)$}{Surface pressure frequency PSD at the trailing edge.}%
where $\delta^*$ is the boundary layer displacement thickness, $U_X$ the chordwise flow velocity and $\overline{\omega} = \omega \delta^* / U_X$,
\begin{align}
  \label{eq:delta_star}
  \delta^* &= \left\{
  \begin{aligned}
    & C(24.3 + 0.6625 \chi) 10^{-4},  \quad \text{if} \quad \chi \leq 4^\circ \\
    & C(26.95 + 0.6625(\chi - 4) + 0.3044(\chi - 4)^2 + 0.0104(\chi - 4)^3) 10^{-4},  \quad 
\chi > 4^\circ,
  \end{aligned}\right.
\end{align}
where $\chi$ is the angle of attack and
\begin{align}
  \label{eq:F}
  F(\overline{\omega}) &= \left\{
  \begin{aligned}
    & \frac{1.732 \times 10^{-3} \overline{\omega}}{1 - 5.489 \overline{\omega} + 36.74 \overline{\omega} ^2 + 0.1505 \overline{\omega} ^5} \quad \text{if} \quad \overline{\omega} < 0.06, \\
    & \frac{1.4216 \times 10^{-3} \overline{\omega} }{0.3261 + 4.1837 \overline{\omega} + 22.818 \overline{\omega}^2 + 0.0013 \overline{\omega} ^3 + 0.0028 \overline{\omega} ^5} \quad \text{if} \quad \overline{\omega} \geq 0.06.
\\
  \end{aligned}\right.
\end{align}
\subsection{Discussion}
\label{sec:discussion}
A number of approximations were made in this section to simplify the comparison between the formulation of section~\ref{cha:kim-george-approach} and that of Amiet (section~\ref{sec:amiets-appr-rotat}). These approximations may need to be replaced with more accurate models for practical applications. 

The first approximation is the equation~\eqref{eq:ly}, giving the correlation length $l_S$, that was derived by Corcos~(\citeyear{corcos1964structure}) for a turbulent boundary layer over a flat plate with zero pressure gradient. A more general equation has been proposed by~\citet{Roger2005}.

A second approximation is the spanwise dependence of the acoustic lift described in figure~\ref{fig:acoustic_lift_ky}. This approximation was made for convenience and does not affect the rest of the analysis. 

Thirdly, the wall-pressure spectrum of Chou and George~(\citeyear{Chou1984}) in section~\ref{sec:trailing-edge-noise}(\ref{sec:surf-press-power}) may be replaced by more complex models such as that of~\citet{Rozenberg2010}. Rozenberg's model takes into account the influence of an adverse pressure gradient and should therefore better predict the effect of aerofoil camber and thickness. However, more research is needed in this area and, ideally, the wall-pressure spectrum should be measured experimentally. 

\section{Schlinker and Amiet's approach for rotating blades}
\label{sec:amiets-appr-rotat}
This section is a detailed and corrected derivation of the theory presented by~\cite{Schlinker1981a}. In particular, the derivation of the present source position is both simpler and more general. This section explains how to apply a theory derived in the wind tunnel reference frame, where a stationary source and observer are immersed in a uniform flow (see section~\ref{sec:trailing-edge-noise}), to the general case, where both the source and the fluid are moving relative to the observer. 

\cite{Amiet1986} observed that the sound emitted by a rotating blade is approximately equal to that emitted by a translating blade at the same location.
His observation was based on the expression of~\cite{Lowson1965} for the pressure radiating from a rotating dipole: the difference between a rotating blade and a translating blade lies in the acceleration of the source in the direction of the observer. 
Amiet argued that when the angular velocity $\Omega$ is much larger than the source frequency $\omega'$, the acceleration term becomes negligible. 
This high frequency assumption will be validated in section~\ref{sec:results} against the new formulation of section~\ref{cha:kim-george-approach}. 
In the following, we will assume that the blade is in rectilinear motion. 

In the situation we are considering here, the observer is stationary relative to the hub and the blade moves through the fluid (figure~\ref{fig:rotor_reference_frame}). 
However the formulae of section~\ref{sec:trailing-edge-noise}, for a blade in a uniform flow, assume that the observer is stationary relative to the blade. 
We therefore examine the current problem successively in these two reference frames. 

\begin{figure}[hbtp]
\centering
  \subfigure[Reference frame of the observer (O): stationary observer, blade moving in uniform flow.]{\includegraphics[width=\textwidth]{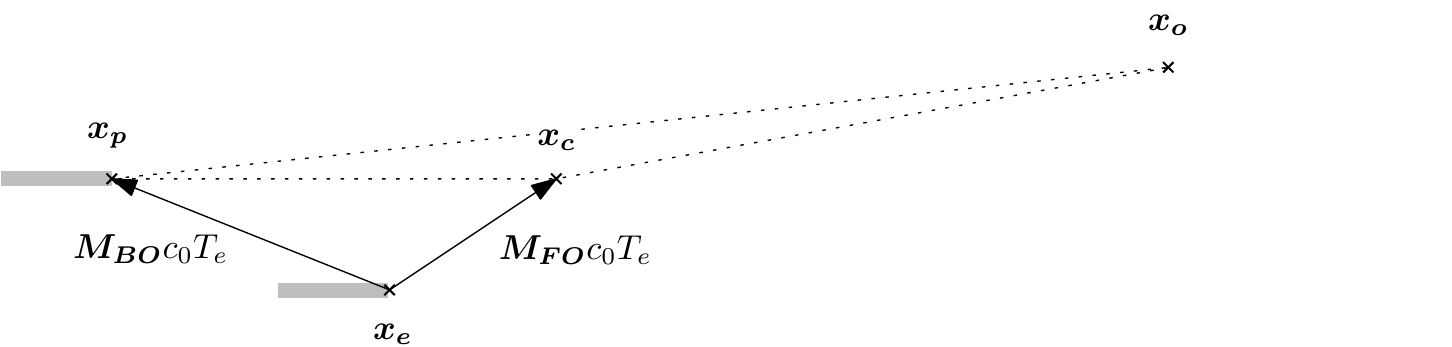}}\\
  \subfigure[Reference frame of the blade (B): equivalent to isolated aerofoil case of figure~\ref{fig:isolated_aerofoil_frame} if $\mbf{M_{OB}} = 0$.]{\includegraphics[width=\textwidth]{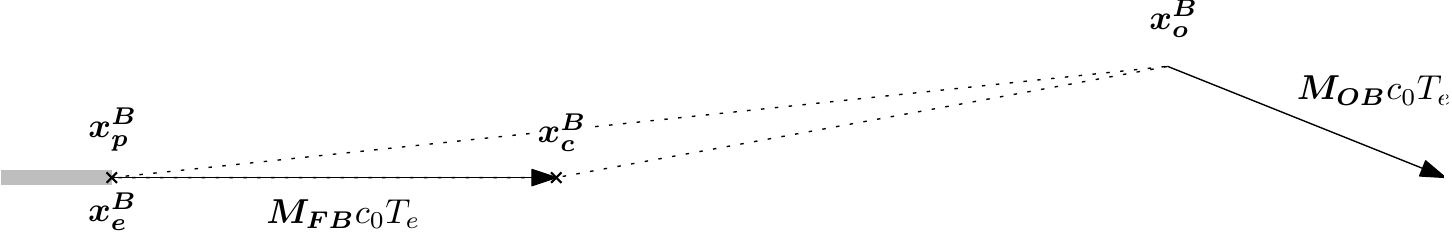}}\\
  \subfigure[Reference frame of the fluid (F): stationary fluid so $|\mbf{x_o^F} - \mbf{x_c^F}| = |\mbf{x_o^F} - \mbf{x_e^F}|= c_0 T_e.$]{\includegraphics[width=\textwidth]{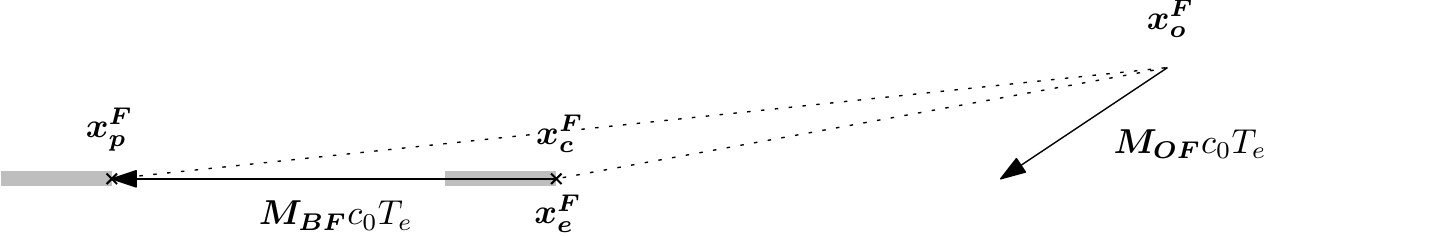}}%
  \caption{Noise radiation from a blade in uniform rectilinear motion analysed in three different reference frames that are fixed to: (a) the observer (O); (b) the blade (B); (c)the fluid (F). $\mbf{M_{AB}}$~stands for ``Mach number of $A$ relative to $B$'', where $A$ and $B$ correspond to one of the reference frames. All figures use the same scale so Mach numbers are comparable. In each case, a pulse of sound is emitted from the emission position $\mbf{x_e}$ at emission time and reaches the observer $\mbf{x_o}$ at reception time, while the source moves with the blade to the present position $\mbf{x_p}$. The convected source position $\mbf{x_c}$ is obtained by convecting the emission position at the Mach number of the fluid for the duration of propagation~$T_e$. The triangle between present source position, convected source position and observer position, shown in dotted lines, is independent of the reference frame: the vectors $\mbf{x_o}-\mbf{x_p}$, $\mbf{x_o}-\mbf{x_c}$ and $\mbf{x_p}-\mbf{x_c}$ are invariant. The position of the blade is shown in grey at emission time and reception time, assuming for simplicity that $\mbf{M_{FB}}$ is aligned with the blade chord (zero angle of attack).}
  \label{fig:reference_frames}
\end{figure}

In the reference frame of the observer ($O$), the observer is stationary and the blade moves rectilinearly at Mach $\mbf{M_{BO}}$ in a uniform flow of Mach $\mbf{M_{FO}}$.
Consider a pulse of sound emitted at $\mbf{x_e}$ from the blade.
When this pulse is received at the observer position $\mbf{x_o}$, the blade has moved to the present source position $\mbf{x_p}$, as illustrated in figure~\ref{fig:reference_frames}(a). 
 
In the reference frame of the blade ($B$), the blade is stationary and the source is fixed to the present source position ($\mbf{x_e^B}=\mbf{x_p^B}$). The Mach number of the flow is $\mbf{M_{FB}} = \mbf{M_{FO}} - \mbf{M_{BO}}$ and the observer is moving at Mach~$\mbf{M_{OB}}$ (figure~\ref{fig:reference_frames}(b)). Comparing figure~\ref{fig:reference_frames}(b) with figure~\ref{fig:isolated_aerofoil_frame}, this situation is equivalent to that of an isolated aerofoil in a uniform flow provided that the observer is stationary, i.e. $\mbf{M_{OB}} = 0$. Here, the observer is moving relative to the blade which leads to a Doppler shift whose impact on the instantaneous PSD will be discussed later.

The above analysis indicates the steps required to apply the formulae derived for an aerofoil in a uniform flow to the present problem:
\begin{itemize}
\item move the origin of the coordinate system from $\mbf{x_e}$ to the present source position $\mbf{x_p}$;
\item rotate the axes of the coordinate system if necessary; 
\item account for the movement of the observer relative to the blade (Doppler shift).
\end{itemize}
\subsubsection{Present source position}

From figure~\ref{fig:reference_frames}(a),
\begin{equation}
  \label{eq:present_position1}
  \mbf{x_p} = \mbf{x_e} + \mbf{M_{BO}} c_0 T_e,
\end{equation}
\nomenclature{$\mbf{x_p}$}{Present source position}%
where $T_e$ denotes the propagation time. 
The above result is more general than the equation (46) in~\cite{Schlinker1981a} that is valid only if the chordwise Mach number of the flow equals $\mbf{M_{FB}}$, i.e. if the angle of attack equals zero.
The source position at emission time satisfies $|\mbf{x_e}| \leq R_t$, where $R_t$ is the blade tip radius, while $T_e \gg 1$ for propagation to the far-field, therefore (provided~$\mbf{M_{BO}} \neq
0$)
\begin{equation}
  \label{eq:present_position2}
  \mbf{x_p} \approx \mbf{M_{BO}} c_0 T_e. \quad (\text{far field}).
\end{equation}
For an observer in the far field, we can hence assume $\mbf{x_e} = 0$ so that the source is located at the hub. 
\subsubsection{Propagation time}
As demonstrated in the following, the propagation time $T_e$ is related to the convected source position $\mbf{x_c}$.
The convected source position is defined as the emission position $\mbf{x_e^F}$ in the reference frame of the fluid (figure~\ref{fig:reference_frames}(c)). 
In that reference frame, the fluid is stationary so the wavefronts are spherical and acoustic waves are propagating at the speed of sound. 
This allows us to relate the propagation time $T_e$ to $\mbf{x_o}$ and $\mbf{x_c}$ since
\begin{equation}
  \label{eq:Te0}
  c_0 T_e \equiv |\mbf{x_o^F} - \mbf{x_e^F}| = |\mbf{x_o^F} - \mbf{x_c^F}|=|\mbf{x_o} - \mbf{x_c}|, 
\end{equation}
where we have used the fact that $\mbf{x_o} - \mbf{x_c}$ is independent of the reference frame (see figure~\ref{fig:reference_frames}).

In general the convected source position $\mbf{x_c}$ is obtained by seeding the flow with a small particle that is released from the source position at emission time and convects at the Mach number of the ambient fluid; $\mbf{x_c}$ corresponds to the position of the particle at reception time.  Using the reference frame of the observer (figure~\ref{fig:reference_frames}(a)), we get
\begin{equation}
  \label{eq:present_position0a}
  \mbf{x_c} = \mbf {x_e} + \mbf{M_{FO}} c_{0} T_e \approx \mbf{M_{FO}} c_{0} T_e, \quad (\text{far field}).
\end{equation}

Substituting~\eqref{eq:present_position0a} into~\eqref{eq:Te0} and taking the square of the result, 
\begin{equation}
  \label{eq:Te1}
  (c_0 T_e)^2 \approx |\mbf{x_o} - \mbf{M_{FO}} c_0 T_e|^2, \quad (\text{far field}).
\end{equation}
The above equation leads to a second order polynomial equation in $R_e \equiv c_0 T_e$ whose solution is given by
\begin{equation}
  \label{eq:Te}
  R_e = \frac{R\left(-M_{FO} \cos \Theta + \sqrt{1 - M_{FO}^2 \sin^2 \Theta}\right)}{1-M_{FO}^2}, \quad (\text{far field}), 
\end{equation}
where $\Theta$ denotes the angle between $\mbf{M_{FO}}$ and $\mbf{x_o}$.
\subsubsection{Coordinate system rotation}
\label{sec:coord-syst-transf}
\begin{figure}[hbtp]
  \centering
  \subfigure[$(x_2, y_2, z_2)$  is obtained by rotating the $(x_1, y_1, z_1)$ by $\pi/2 - \gamma$ around the $z_1$-axis.]{\includegraphics[height=4cm]{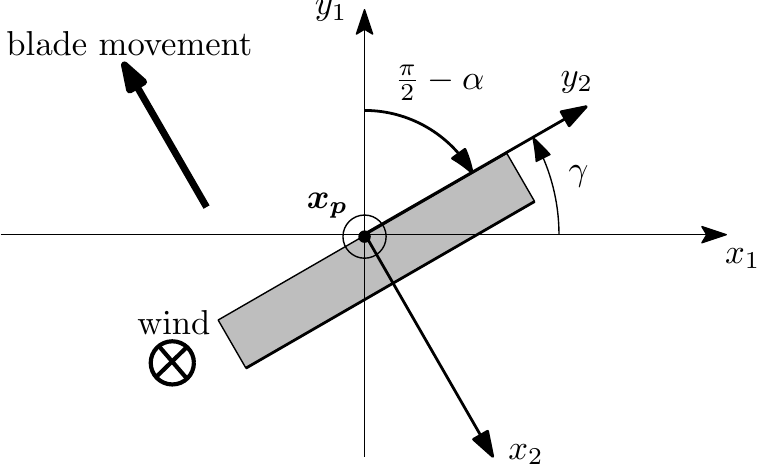}} \quad
  \subfigure[$(X, Y, Z)$ is obtained by rotating $(x_2, y_2, z_2)$ by $\alpha$ around the $y_2$-axis.]{\includegraphics[height=4cm]{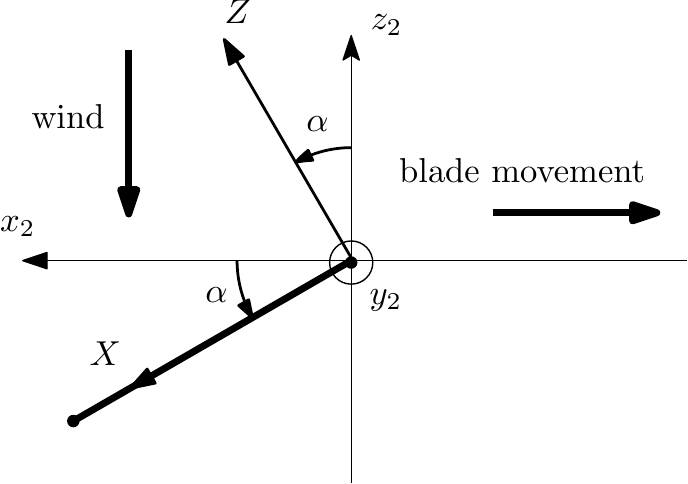}}
  \caption{Coordinate systems $(x_1, y_1, z_1)$, centred on the present source position $\mbf{x_p}$, $(x_2, y_2, z_2)$, such that $y_2$ is in the spanwise direction, and $(X, Y, Z)$, such that $Y$ is in the spanwise direction and $X$ in the chordwise direction.} 
  \label{fig:x1_to_x2}
\end{figure}
In the reference frame of the blade used in section~\ref{sec:trailing-edge-noise}, the $y$-axis is pointing spanwise and the $x$-axis is pointing chordwise. 
Let $\mbf{x_1} = (x_1, y_1, z_1)$ denote the coordinate system centred on the present source position ($\mbf{x_1} = \mbf{x_o} - \mbf{x_p}$). 
For the $y$-axis to point in the spanwise direction, the $\mbf{x_1}$-coordinate system is rotated by $\pi/2 - \gamma$ around the $z_1$-axis (figure \ref{fig:x1_to_x2}(a)), i.e.
\begin{equation}
  \label{eq:x2}
  \mbf{x_2} = \underline{\mbf{R_z}} (\pi/2 - \gamma) \mbf{x_1},
\end{equation}
where $\underline{\mbf{R_z}}$ denotes the rotation matrix about the $z_1$-axis (see equation~\eqref{eq:rotation_matrices}).

Similarly, for the $x$-axis to point in the chordwise direction, the $\mbf{x_2}$-coordinate system is rotated by $\alpha$ around the $y_2$-axis (figure \ref{fig:x1_to_x2}(b)):
\begin{equation}
  \label{eq:x3}
  \mbf{X} = \underline{\mbf{R_y}} (\alpha) \mbf{x_2},
\end{equation}
where $\underline{\mbf{R_y}}$ denotes the rotation matrix about the $y_2$-axis (see equation~\eqref{eq:rotation_matrices}).

The transform from the observer reference frame (figure~\ref{fig:reference_frames}(a)) to the blade reference frame is hence given by
\begin{equation}
  \label{eq:tranform}
  \mbf{X} = \underline{\mbf{R_y}}(\alpha) \underline{\mbf{R_z}}(\pi/2 - \alpha) (\mbf{x_o} - \mbf{x_p}).
\end{equation}
\subsubsection{Doppler shift}
\label{sec:doppler-shift}
The sound frequency $\omega$ at the observer is shifted compared to the emission frequency $\omega'$. 
The ratio $\omega/\omega'$ is called the Doppler shift and is a function of the source Mach number $\mbf{M_{BO}}$ and the flow Mach number $\mbf{M_{FO}}$. 
As explained by~\cite{Amiet1974}, the Doppler shift is given by
\begin{equation}
  \label{eq:period_observer_approx}
  \frac{\omega}{\omega'} =  1 + \frac{\mbf{M_{BO}} \cdot \mbf{\widehat{CO}}}{1 +  (\mbf{M_{FO}} - \mbf{M_{BO}}) \cdot \mbf{\widehat{CO}}} \quad \text{(far field)},
\end{equation}
\nomenclature{$\omega'$}{Angular frequency of the source}%
where $\mbf{\widehat{CO}}$ is the unit vector from the convected source position ($\mbf{x_c}$ in figure~\ref{fig:reference_frames}) to the observer position. 
\label{sec:power-spectr-dens}

It can be shown (\cite{Amiet1974} and appendix~\ref{sec:inst-freq-doppl}) that in the far field the PSD for a moving observer, $S_{pp}$, is related to the PSD for a fixed observer, $S_{pp}'$ by $S_{pp}(\omega) = (\omega' / \omega) S_{pp}'(\omega')$, therefore
\begin{equation}
  \label{eq:instant_Spp}
  S_{pp}(\mbf{x_o}, t, \omega) = \frac{\omega'}{\omega} S_{pp}'(\mbf{X}, \tau, \omega'),
\end{equation}
where $S_{pp}'$ can be estimated using the isolated aerofoil expression of~\eqref{eq:psd_isolated}.
Furthermore, equation~\eqref{eq:psd_isolated} is expressed in terms of the chordwise Mach number ${M_X = \mbf{M_{FB}} \cdot \mbf{\hat X}}$ of the flow relative to the blade. 
A derivation of $M_X$, extending that of~\cite{Schlinker1981a} to the case of a non-zero angle of attack, is given in appendix~\ref{sec:chordw-mach-numb}.

The time averaged PSD is obtained by averaging the instantaneous PSD, i.e. equation~\eqref{eq:instant_Spp}, over one rotation of the rotor. Since the instantaneous PSD is a function of source time $\tau$, the time increment $dt$ must be expressed in terms of $d\tau$. If $n$ is the number of acoustic periods measured at the observer location during $dt$ then $dt = n 2\pi /\omega$.
The time taken for the source to generate those $n$ periods is $d\tau = n 2\pi / \omega'$ so $dt = (\omega' / \omega) d\tau$. Finally, it is convenient to express $d\tau$ in terms of the blade azimuthal angle $\gamma$ using the relation $\Omega = d\gamma / d\tau$. The time averaged PSD from a rotating blade element is hence given by

\begin{equation}
  \label{eq:amiet_Spp_end}
  S_{pp}(\mbf{x_o}, \omega) = \frac{1}{2 \pi} \intx{0}{2\pi}{ \left(\frac{\omega'}{\omega}\right)^{2} S_{pp}'(\mbf{X}, \omega', \gamma) }{\gamma}.
\end{equation}
The above result agrees with~\cite{Schlinker1981a}, provided that $\omega$ is replaced by $\omega_o$ in the left hand side of (54) in their report, and $\omega$ and $\omega_o$ are swapped in the left hand side of their equation (56).
Note that there has been some discrepancy in the literature about the exponent in the Doppler term of~\eqref{eq:amiet_Spp_end}. 
For example, \cite{Amiet1977} initially proposed a value of 1 for the exponent. The same is considered by \cite{Rozenberg2010}.
This ignores the additional weighting of the time increment and is equivalent to averaging over the angular position of a blade segment.
We propose that the correct value of the exponent is 2. In section~\ref{sec:results}, evidence for this is provided by comparison of predictions using the two methods presented in sections~\ref{cha:kim-george-approach} to \ref{sec:amiets-appr-rotat}.

\section{Results}					  
\label{sec:results}
In this section, both Amiet's formulation  (equation~\eqref{eq:amiet_Spp_end}) and our new formulation (equation~\eqref{eq:Spp_final_simple}) are applied to several model blade elements. These blade elements have been chosen to cover the range of applications featured in the literature on trailing edge noise: cooling fans~\citep{Rozenberg2010}, open propellers~\citep{Pagano2010}  and wind turbines\citep{Glegg1987, Oerlemans2010}. Each blade element is described through its distance from the hub (radius), its geometry (chord and pitch angle $\alpha$), and the set of Mach numbers (blade Mach $M_{BO}$, flow Mach numbers relative to the observer $M_z = M_{FO}$ and to the blade $M_X=M_{FB}$), whose values have been chosen to match a typical situation. For example, a typical wind turbine element has a large chord, low pitch angle and rotates at low Mach number in a low wind speed. The specific values chosen here, presented in table~\ref{tab:params}, were the ones used by \citet{Blandeau2011a}. 
The span is arbitrarily equal to the chord.
Note that for each blade element, the values are not independent; the three Mach number form a velocity triangle wherein the pitch angle is between $\bm{M_{FO}}$ and $\bm{M_{FB}}$. Furthermore, the values would usually change along a full blade. 
\begin{table}[hbt]
  \centering
\begin{tabular}{llll}
 & Cooling fan & Wind turbine & Open propeller \\
\hline
radius & 0.40 m & 29 m & 1.80 m \\
chord & 0.13 m & $\phantom{0}$2 m & 0.31 m \\
$M_{BO}$ & 0.0525 & 0.165 & 0.748 \\
$\alpha$ & 34 $\deg$ & 10 $\deg$ & 13 $\deg$ (take-off), 38 $\deg$ (cruise) \\
$M_{FO}$ & 0.0354 & 0.029 & 0.584 (take-off), 0.228 (cruise) \\
$M_{FB}$ & 0.0633 & 0.167 & 0.949 (take-off), 0.782 (cruise) \\
\end{tabular}
\caption{Typical parameters for the blade segments used in three different applications: open-propeller, cooling fan and wind turbine. 
These parameters are the ones proposed by~\cite{Blandeau2011a}. The span is defined as a third of the radius. 
The wind Mach numbers, relative to the observer  $M_{FO}$(${=M_z}$) or to the blade $M_{FB}$(${=M_X}$) can be obtained from the pitch angle $\alpha$ and blade speed $M_{BO}$ but are given for completeness.}
\label{tab:params}
\end{table}

Although this paper focuses on noise radiation from small blade elements, the noise spectrum of a full blade may be obtained using strip theory. In strip theory, one divides the blade into small elements along its radius and sums the (uncorrelated) noise spectra generated from each element. Limitations and extensions of strip theory are discussed in~\citet{christophe2009amiet}. In addition to the approximations discussed in section~\ref{sec:trailing-edge-noise}(\ref{sec:discussion}), the thin-blade approximation has been used. This is not a good approximation for angles close to the propeller axis where a numerical approach taking the thickness of the blade into account may be preferable~\citep{zhou2006frequency}.
\begin{figure}[hbt]
\centering
\subfigure[Wind turbine]{\includegraphics[height=3.5cm]{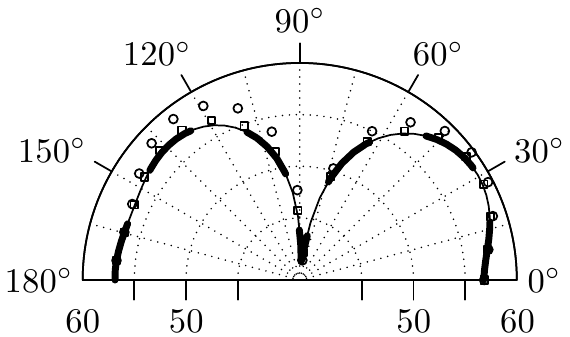}}
\subfigure[Open propeller at takeoff.
]{\includegraphics[height=3.5cm]{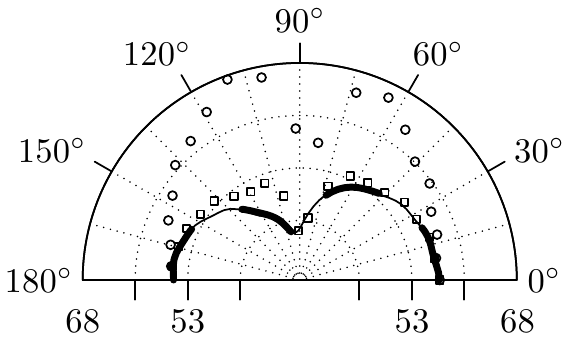}}
\caption{Sound pressure level directivity using the new formulation (equation~\eqref{eq:Spp_final_simple}, dashed line) and Amiet's formulation (equation~\eqref{eq:amiet_Spp_end}) with the exponents of the Doppler shift set to $-2$ ($\circ$ symbols), $1$ ($\small \square$ symbols) and $2$ (solid line).
The frequency is $kC=5$. The decibel scale assumes $|\mathbf{x_o}| = 1\, \text{m}$ and wind speed is from right to left.}
\label{fig:exponents}
\end{figure}

Figure~\ref{fig:exponents} compares the sound pressure level directivity obtained using the new formulation and Sclinker and Amiet's approach, when the exponents of the Doppler shift (see equation~\eqref{eq:amiet_Spp_end}) takes the values 1, 2 and -2 taken from the literature.
A low Mach number case (wind turbine) is shown in figure~\ref{fig:exponents}(a), and a high Mach number case (open propeller) in figure~\ref{fig:exponents}(b).
Differences of up to $5$ dB  and $20$ dB can be seen when Schlinker and Amiet's approach are used with an exponent of $1$ (square symbols) and $-2$ (round symbols) respectively. The best agreement between the new approach and Schlinker and Amiet's approach is obtained for an exponent of $2$ ($+$ symbols), which validates equation~\eqref{eq:amiet_Spp_end}.

\begin{figure}[phbt]
\centering
\subfigure[Cooling fan]{
\label{fig:3}
    \includegraphics[width=.29\textwidth]{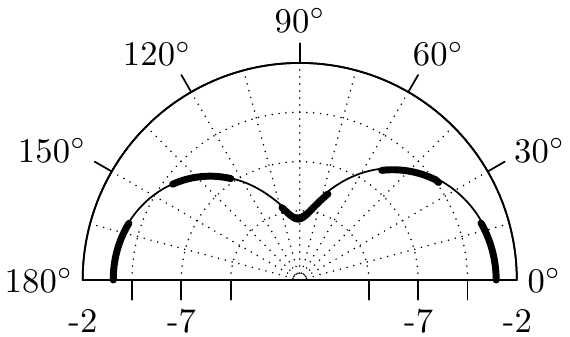} \quad
    \includegraphics[width=.29\textwidth]{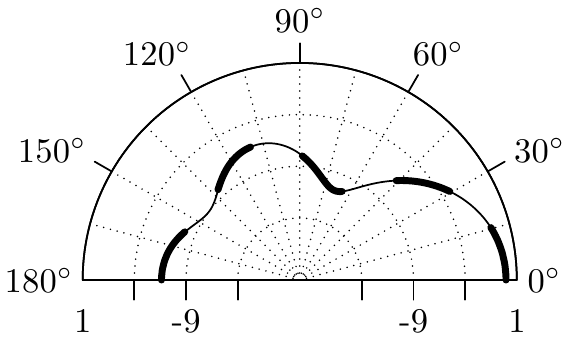} \quad
    \includegraphics[width=.29\textwidth]{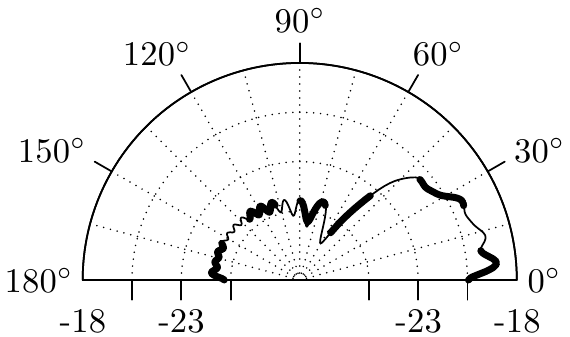}
}
\subfigure[Wind turbine]{
\label{fig:4}
{\includegraphics[width=.29\textwidth]{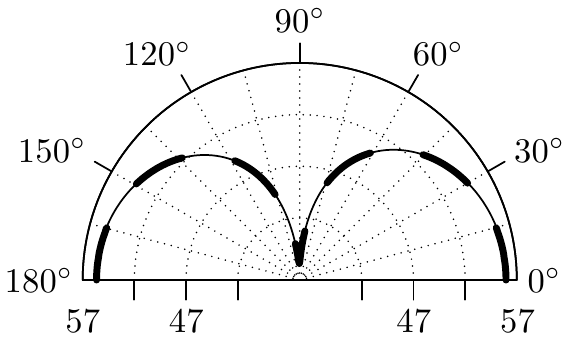}} \quad
{\includegraphics[width=.29\textwidth]{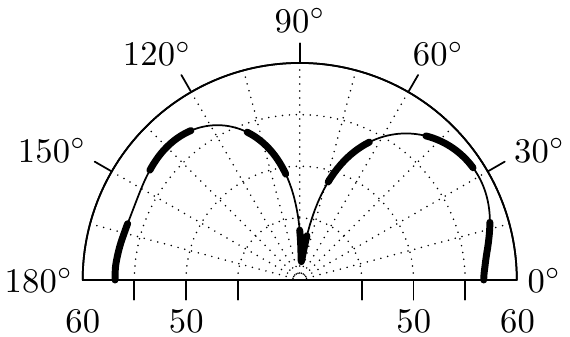}} \quad
{\includegraphics[width=.29\textwidth]{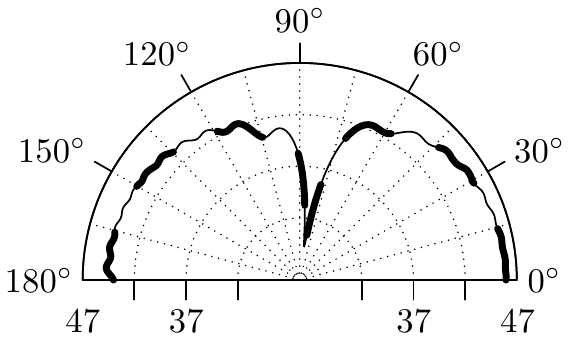}} %
}
\subfigure[Open propeller at take-off ]{
\label{fig:1}
{\includegraphics[width=.29\textwidth]{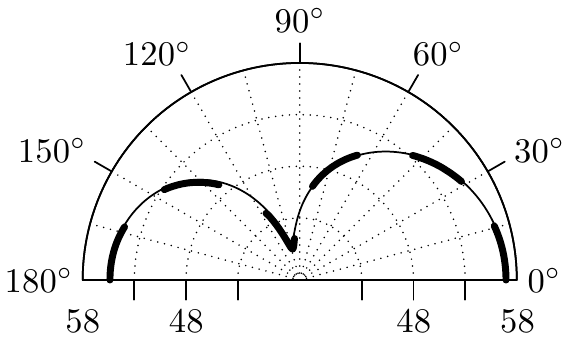}}\quad
{\includegraphics[width=.29\textwidth]{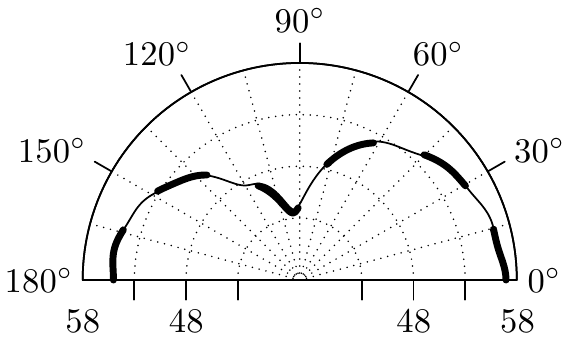}}\quad
{\includegraphics[width=.29\textwidth]{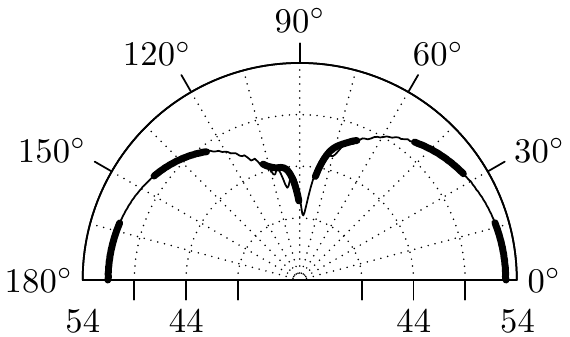}}%
}
\subfigure[Open propeller at cruise]{
\label{fig:2}
{\includegraphics[width=.29\textwidth]{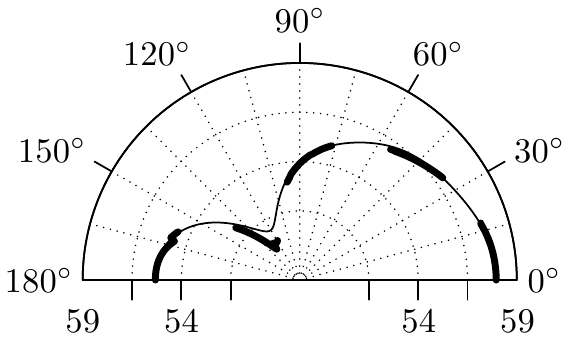}}\quad
{\includegraphics[width=.29\textwidth]{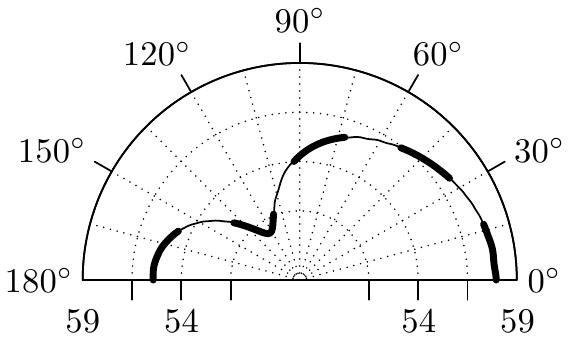}}\quad
{\includegraphics[width=.29\textwidth]{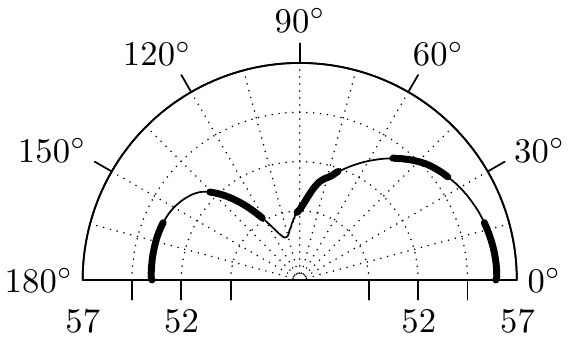}}%
}
\caption{Sound pressure level directivity in decibels for the model blade elements of table~\ref{tab:params}, and an observer $\mbf{x_o}$ such that $|\mbf{x_o}|=1\,\text{m}$ (far field is implied) and $\gamma_o = 0^{\circ}$. The normalized frequency $kC$ equals $0.5$ (left column), $5$ (middle) or $50$ (right). Solid line is Amiet's approach (equation~\eqref{eq:amiet_Spp_end}), dashed line is the new formulation of equation~\eqref{eq:Spp_final_simple}. Wind speed is from right to left.}
\label{fig:results}
\end{figure}

Figure~\ref{fig:results} shows the sound pressure level directivity for the blade segments defined in table~\ref{tab:params} that model a cooling fan (figure \ref{fig:3}), a wind turbine (figure \ref{fig:4}), and an open propeller at take-off (figure \ref{fig:1}) and cruise (figure \ref{fig:2}).
In each case, the results are obtained using the new formulation (dashed line) and Amiet's approach (solid line). In figures~\ref{fig:exponents} and \ref{fig:results} the levels are normalised to give an effective observer distance of $|\mbf{x_o}|=1\,\text{m}$ and the observer is located at $\gamma_o = 0^{\circ}$.

The two approaches agree to within 1~dB in all cases. These results suggest that Schlinker and Amiet's theory is valid when the acceleration effects are negligible, i.e. $\omega \gg \Omega$, and when Amiet's blade response function is applicable, i.e. $kC > 1$. Since $\omega/\Omega = k r / M_{BO}$, this ratio is larger than $2.9$ for all the test cases in table \ref{tab:params}. Note also that Schlinker and Amiet's formulation is applicable even for high speed open rotors: figure~\ref{fig:2} shows a good agreement for the open propeller model at cruise for which the chordwise Mach number is very high ($M_{FB} = 0.95$).

\section*{Conclusions}
Schlinker and Amiet's~(\citeyear{Schlinker1981a}) theory of trailing
noise from rotating blade has been validated by comparison with a new formulation that implicitly includes acceleration effects. The results presented in this paper show that
the modified form of Schlinker and Amiet's predicts sound pressure levels within 1~dB of our new formulation for subsonic chordwise Mach numbers of up to 0.95 at high frequencies ($kC > 1$ and $\omega \gg \Omega$). 
It is therefore applicable to both low speed applications (cooling fans,
wind turbines) and high speed ones (propellers).

We propose that the correct exponent for the Doppler weighting $(\omega' / \omega)^{a}$ in Schlinker and Amiet's theory is $a=2$.
Using $a=1$, as in~\cite{Amiet1977} and~\cite{Rozenberg2010} gives results that agree with our formulation to within 5~dB.
However, all other things being equal, putting $a=-2$ in Schlinker and Amiet's theory  overestimates the result by up to 20~dB at high Mach numbers. 

The range of validity of Amiet's theory for rotating blades is currently limited to high frequencies for two reasons: it uses a high frequency blade response function and neglects the effect of blade acceleration. More research is needed to identify the most important effect between the two. If the effect of acceleration can be neglected, i.e. at high frequency relative to the rotational speed, Amiet's theory can be extended to lower frequencies by using a low frequency blade response function (\cite{Roger2005}). 

\ack{Samuel Sinayoko and Anurag Agarwal wish to acknowledge the support of
Mitsubishi Heavy Industries.
Mike Kingan wishes to acknowledge the
continuing financial support provided by Rolls-Royce plc.
through the
University Technology Centre in Gas Turbine Noise at the Institute of
Sound and Vibration Research.
The authors wish to thank Vincent
Blandeau, Phil Joseph, Thomas Nod\'{e}-Langlois, Stephane Moreau, Michel Roger, Chris Morfey and Ann
Dowling for their thoughtful comments.} 


\appendix
\section*{Appendix}					  
\section{Far field Green's function}
\label{sec:far-field-greens}
Taking the Fourier transform of \eqref{eq:garrick_watkins_notations} over $t$, and using the far field approximation of~\eqref{eq:garrick_watkins_notations} yields
\begin{equation}
  \label{eq:G_garrick_watkins_fourier}
  G(\mbf{x}, \tau | \mbf{x_o}, \omega) = \frac{e^{-i\omega(\tau + s / c_0)}}{4\pi s} = \frac{e^{-i\omega\tau}}{4\pi s} e^{i(k_z z + k_r r \cos(\gamma - \gamma_0))},
\end{equation}
where $k_z = k_D \cos \theta_e$ and $k_r = k_D \sin \theta_e$. Using the Jacobi-Anger expansion
\begin{equation}
  \label{eq:Jacobi_Anger}
  e^{i k_r r \cos(\gamma - \gamma_0)} = \sum_{n=-\infty}^{n=+\infty} J_n(k_r r) e^{i n (\gamma - \gamma_0 + \pi/2)},
\end{equation}
which yields equation~\eqref{eq:Gw}.

\section{Rotation matrices}					  
\begin{align}
  \label{eq:rotation_matrices}
\underline{\mbf{R_z}}(\theta) &= %
\begin{pmatrix}
  \cos \theta & -\sin \theta & 0 \\
  \sin \theta & \cos \theta & 0 \\
  0 & 0 & 1 
\end{pmatrix}
,&%
\underline{\mbf{R_y}}(\theta) &= %
\begin{pmatrix}
  \cos \theta & 0 & -\sin \theta \\
  0 & 1 & 0 \\
  \sin \theta & 0 & \cos \theta 
\end{pmatrix}
.
\end{align}

\section{Instantaneous frequency and Doppler shift}
\label{sec:inst-freq-doppl}
\subsection{Physical interpretation relative to the speed of time}
\label{sec:phys-interpr-relat}
The Doppler shift indicates how the frequency of a pure tone varies for a moving observer (relative to the source), compared to a fixed observer.
Following~\citet{Amiet1974}, an alternative physical interpretation  can be obtained by observing that, for a pure tone, the pressure $p'$ for a fixed observer and $p$ for a moving observer are given by
\begin{equation}
  \label{eq:doppler_interpretation}
  \left\{
  \begin{aligned}
  p'(t) &= e^{-i\omega' t}  \\
  p(t) &= e^{-i \omega t}  
  \end{aligned}\right.
  , \quad \text{so} \quad p(t) = e^{-i \omega'(\omega/\omega')t} = p'\left( (\omega / \omega') t \right),
\end{equation}
where $\omega'$ is the source frequency and $\omega$ the frequency received by the moving observer.
The above equation implies that the measurement of the pressure time history is sped up (or slowed down), by a factor $\omega / \omega'$, for a moving observer, compared to a fixed observer.
\subsection{Power spectral density in two reference frames}
\label{sec:power-spectr-dens-1}
From equation~\eqref{eq:doppler_interpretation}, the autocorrelations $R_{pp}'(t)$ and $R_{pp}(t)$ for a fixed observer and a moving observer (relative to the source), are such that
\begin{align}
  \label{eq:rpp}
  R_{pp}(t) &= \lim_{T\to+\infty} \frac{1}{2T}\intx{-T}{T}{p(\tau) p(t - \tau)}{\tau} = \lim_{T\to+\infty} \frac{1}{2T}\intx{-T}{T}{p'(a \tau) p'(a (t - \tau))}{\tau} \\
           &= \lim_{T\to+\infty}\frac{1}{2aT}\intx{-aT}{aT}{p'(\tau') p'(a t - \tau')}{\tau'} = R_{pp}'(at),
\end{align}
\nomenclature{$R_{pp}(t)$}{Autocorrelation function of the pressure at the observer location}%
where $a = \omega / \omega'$.

Since the power spectral density (PSD) is defined as the Fourier transform of the autocorrelation, the PSD $S_{pp}(\omega)$, for a moving observer, is related to the PSD $S_{pp}'(\omega')$ for a fixed observer by
\begin{equation}
  \label{eq:Spp_calculation}
  S_{pp}(\omega) = \intxInf{R_{pp}(t) e^{i\omega t}}{t} = \intxInf{R_{pp}'(a t) e^{i \omega' a t}}{t} = \frac{1}{a} S_{pp}'(\omega'),
\end{equation}
\nomenclature{$S_{pp}'(\omega)$}{Pressure PSD produced by a stationary aerofoil immersed in a flow and measured by a stationary observer}%
i.e.

\begin{equation}
  \label{eq:Spp}
  S_{pp}(\omega) = \frac{\omega'}{\omega} S_{pp}'(\omega').
\end{equation}
\section{Chordwise Mach number}
\label{sec:chordw-mach-numb}
\begin{figure}[hbtp]
  \centering
  \subfigure[{Flat plate (in grey) in the $(\hat{\boldsymbol{\gamma}}, \mbf{\hat{\mbf{z}}})$ plane, where $\mbf{\hat{\mbf{z}}} $ is pointing against the flow and is orthogonal to the rotor plane, and $\hat{\boldsymbol{\gamma}}$ is in the direction of rotation.%
The angle between the rotor plane and flat plate is $\alpha$ and the angle of attack is $\chi$.%
The flow Mach number relative to the blade is $\mbf{M_{FB}} = -M_z \mbf{\hat{\mbf{z}}} - M_\gamma \hat{\boldsymbol{\gamma}}$.
The chordwise Mach number is $M_{X}$.}]{\includegraphics[width=.45\textwidth]{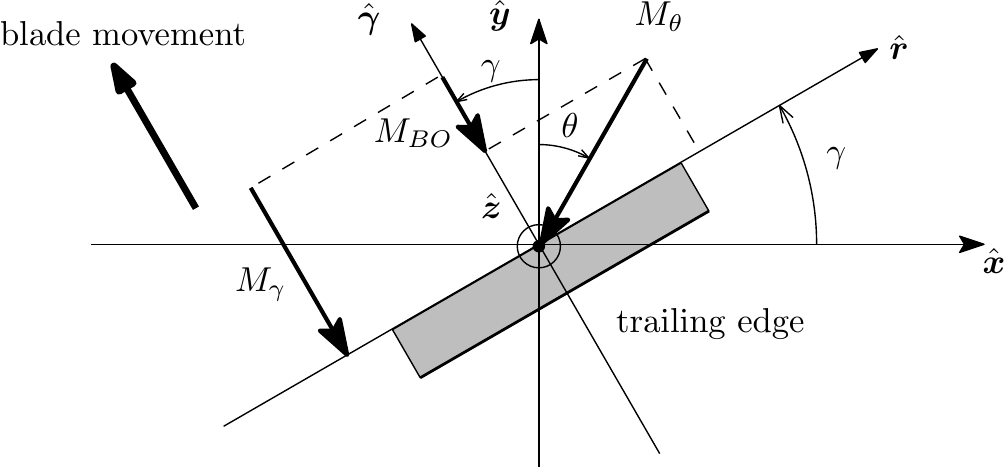}}  \qquad
\label{fig:1_egamma_ez}
  \subfigure[{Relationship between the flow Mach number $M_\gamma$ along $\hat{\boldsymbol{\gamma}}$, the blade Mach number $M_{BO}$ and the (wind) cross-flow $M_\theta$ in the rotor plane $(\mbf{\hat{\mbf{x}}}, \mbf{\hat{\mbf{y}}})$.
Note that the flat plate is not in the rotor plane (see~figure~(a)). 
The angle $\gamma$ is the azimuthal angle of the blade. 
The angle $\theta$ gives the direction of the cross flow relative to $\mbf{\hat{\mbf{y}}} $.}]{\includegraphics[width=.45\textwidth]{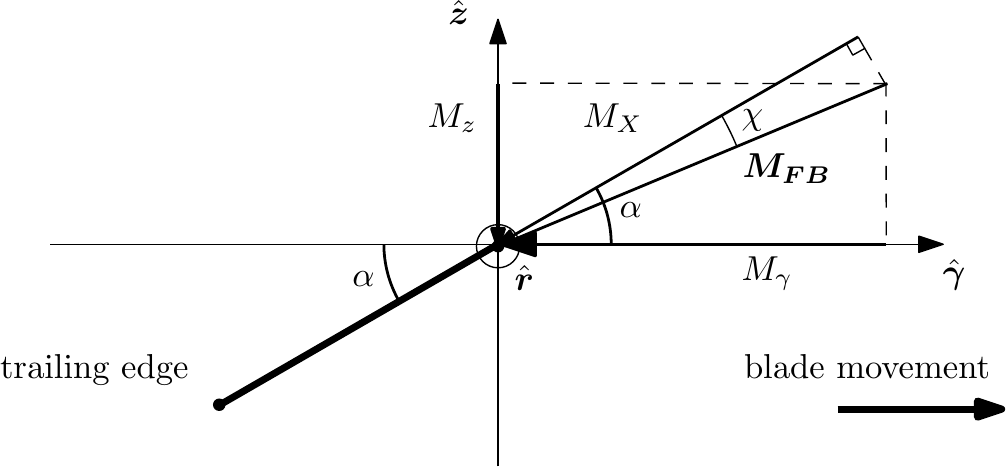}}
\label{fig:2_ex_ey}
\caption{Coordinate systems and Mach numbers around a flat plate moving in a uniform flow.}
\label{fig:coordinates_isolated}
\end{figure}
We derive the chordwise Mach number $M_X$ given the blade Mach number ${\mbf{M_{BO}} \equiv M_{BO} \hat{\boldsymbol{\gamma}}}$ and the flow Mach number $\mbf{M_{FO}}$.
As illustrated in figure~\ref{fig:coordinates_isolated}(b), $\mbf{M_{FO}}$ can be decomposed into a component $M_z$ normal to the rotor plane (inflow), and a component $M_\theta$ parallel to the rotor plane (cross-flow):
\begin{align}
    \label{eq:M_Mf}
    \mbf{M_{FO}} &= -M_z \mbf{\hat{\mbf{z}}} - M_\theta \hat{\boldsymbol{\theta}}.
\end{align}
From figure~\ref{fig:coordinates_isolated}(a), the chordwise Mach number $\mbf{M_{X}}$ is given by
\begin{align}
  \label{eq:1}
  M_{X} &= M_{FB} \cos \chi = \sqrt{M_\gamma^2 + M_z^2} \cos \chi . 
\end{align}
\nomenclature{$M_{X}$}{Chordwise Mach number of the flow}%
\nomenclature{$\widehat{(~)}$}{Unit vector in rotor plane in $(~)$-direction}%
From figure~\ref{fig:coordinates_isolated}(b), the azimuthal Mach number $M_\gamma$ of the flow relative to the blade can be expressed in terms of $M_{BO}$ and $M_\theta$ by 
\begin{align}
  \label{eq:2}
  M_\gamma &= M_{BO} + M_{\theta} \cos(\gamma + \theta). 
\end{align}
Substituting equation~\eqref{eq:2} into equation~\eqref{eq:1},
\begin{align}
  \label{eq:MX}
  M_{X} &= \cos \chi \sqrt{[M_{BO} + M_\theta \cos(\gamma + \theta)]^2 + M_z^2 } . 
\end{align}

\end{document}